# Coherent control of the causal order of entanglement distillation

Zai Zuo ,[*] Michael Hanks, and M. S. Kim

*QOLS, Blackett Laboratory, Imperial College London, SW7 2AZ, United Kingdom*

Indefinite causal order is an evolving field with potential involvement in quantum technologies. Here we propose and study one possible scenario of practical application in quantum communication: a compound entanglement distillation protocol that features two steps of a basic distillation protocol applied in a coherent superposition of two causal orders. This is achieved by using one faulty entangled pair to control-swap two others before a fourth pair is combined with the two swapped ones consecutively. As a result, the protocol distills the four faulty entangled states into one of a higher fidelity. Our protocol has a higher fidelity of distillation and probability of success for some input faulty pairs than conventional concatenations of the basic protocol that follow a definite distillation order. Our proposal shows the advantage of indefinite causal order in an application setting consistent with the requirements of quantum communication.

## I. INTRODUCTION

Causality is a fundamental concept in nature and deeply embedded in the traditional model of computation. A computing algorithm, classical or quantum, usually envisions a target system undergoing a series of gates in a fixed causal order. Recent studies [1] revealed that under the assumption that quantum mechanics is valid locally, events can happen in an indefinite causal order. This concept was then extended to a quantum computation model with an indefinite causal structure, which features quantum operations occurring in an indefinite causal order [2,3].

Significant effort has been made to look for specific communication and computational tasks where indefinite causal structures provide advantage. On the computation side, there are specific classes of problems such as Fourier [4,5] and Hadamard promise problems [6,7] that have been shown to enjoy a reduction in query complexity when quantum gates were queried through an indefinite causal order. The advantage of such a setup in the task of solving the generalized Deutsch's problem has also been explored [8]. On the communication side, some probabilistic communication tasks were found to result in a boost in success probability [9] and reduction in communication complexity [10] when passage orders of information through communication parties were put into a superposition. It was also discovered that putting two noisy channels through a superposition of passage orders reduces the noise of communication and for some specific channels results in complete noise removal [11–15]. The above examples of indefinite causal order all arise under the setup that two events, each described by a completely positive trace nonincreasing (CPNTI) map, are passed through by the target system in an order coherently controlled by another qubit. Mathematically, the advantage of putting CPNTI maps in an indefinite causal order arises from the noncommutation of Kraus operators that make up each quantum map [12,16,17].

A natural question to ask is whether indefinite causal structures are useful in any current existing or near-term quantum information processing tasks. Studies that reduce communication noise by superposing the passage orders of noisy channels [11–15] have assumed the message making a noiseless return to the front of the other channels after passing through and exiting from the end of the other one, which is inconsistent with realistic situations of quantum communication. On the computation side, information-theoretic tasks that were studied in previous works [4–7,9,10] cannot be easily generalized to solving more common computational problems, leaving significant research effort still required to bridge the gap. A recent work [8] experimentally demonstrated solving the generalized Deutsch's problem with an indefinite causal order setup, but such a proposal is only likely to be useful in the long term with fault-tolerant quantum computers that can solve this problem on a large scale with large inputs. Fault-tolerant quantum computers are of immense experimental challenge to build, which prevents such proposals to be used in the short term. Some other examples [18–20] considered applying two quantum teleportation steps in an indefinite causal order by coherently swapping the two involved entangled states to reduce noise on the teleported state caused by imperfection of the entangled states upon their generation. However, as we show in the Appendix of the paper, their proposal does not achieve noise reduction. We also argue that in the more general case where noise occurs during the process of distribution of entanglement, swapping of the entangled states must be carried out remotely. This requires extra entanglement which brings difficulties to their proposal. It therefore remains largely unknown how useful indefinite causal structures are in near-term computation and communication tasks.

In this paper, we study the entanglement distillation of four entangled pairs and show that distillation of entanglement, which is a basic and necessary task in quantum communication, benefits from indefinite causal structure. More specifically, it allows the production of higher-fidelity entangled pairs than merely carrying out distillation steps in a definite causal order. We consider the well-known entanglement distillation protocol proposed by Deutsch *et.al.* [21] (to be referred to as the DEJMPS protocol) which turns two faulty entangled pairs into one of a higher fidelity. When three faulty pairs, $\chi_i$ where $i \in 1, 2, 3$, are subject to such a protocol, $\chi_1$ can be combined with $\chi_2$ before the distillation product is combined with pair $\chi_3$, or that $\chi_1$ can be merged with $\chi_3$ then $\chi_2$. This defines an "order" of entanglement distillation. We describe a protocol where these two orders are put into a coherent superposition and show that like other previously studied cases of indefinite causal order, the advantage in fidelity also comes from nontrivial commutation of Kraus operators of the CP map that describes the entanglement distillation. We then show the practical advantage of our protocol by presenting that for some input faulty entangled states, characterized by amount of mixture of the four Bell states, our protocol results in a higher fidelity and/or success probability than merely putting two distillation steps in a definite causal order. Given the known connection between entanglement distillation protocols and quantum error correction codes [22], we hope this work will stimulate effort into looking for the advantage of indefinite causal structures in quantum error correction.

This paper is organized as follows. In Sec. II, we review the basic principles of recent applications of indefinite causal order in the form of a quantum switch. In Sec. III, we first review two-way entanglement distillation protocols constructed from small error-detecting codes. This includes the DEJMPS protocol and a protocol using three entangled pairs, the latter of which is constructed from three-bit quantum error detecting code. We then present in Sec. IV a naively modified protocol which performs two DEJMPS distillation steps in an superposed causal order. As the circuit of the proposed protocol is quite complicated, we then describe in Sec. V a simplified protocol which we show still simulates two DEJMPS distillation steps in superposed causal orders. In Sec. VI we argue for our protocol by presenting the parameter regions of the input faulty states where our scheme shows an advantage over concatenations of the DEJMPS protocol and the three-pair protocol that follow a definite causal order. Some conclusions are then provided in Sec. VII.

## II. QUANTUM SWITCH AND KRAUS OPERATORS OF COMPLETELY POSITIVE MAPS

A common framework to realize indefinite causal order between two quantum operations on a target state is to use an additional qubit to control the orders of occurrences of operations, such that the different orders are correlated with different basis states of the control qubit. Such a setup is named a "quantum switch" in the literature. Here, we give a short review of the basic principles of proposed applications of the quantum switch. Quantum operations are mathematically described as completely positive (CP) trace-nonincreasing maps. When two quantum operations $\mathcal{M} = \sum_i M_i \rho M_i^\dagger$ and $\mathcal{N} = \sum_j N_j \rho N_j^\dagger$, where $M_i$ and $N_j$ are Kraus operators (a special case is when $\mathcal{M} = M \rho M$ and $\mathcal{N} = N \rho N$ are unitary operators), are put into a quantum switch controlled by a qubit $\rho_c$, the overall map on the control and target states reads

$$\mathcal{D}(\rho_c, \rho) = \sum_{ij} W_{ij}(\rho_c \otimes \rho) W_{ij}^\dagger, \qquad (1)$$

where the overall set of Kraus operators reads $W_{ij} = |0\rangle\langle 0| \otimes M_j N_i + |1\rangle\langle 1| \otimes N_i M_j$.

Previous works that studied using a quantum switch [23] to boost the communication capacity of noisy channels [11–14,24] often initialize the control state as $\rho_c = |+\rangle\langle+|$, an even superposition of passage orders through the two quantum maps giving a resulting state

$$
\begin{aligned}
\mathcal{D}(\rho_c, \rho) = \frac{|+\rangle\langle+|}{2} \otimes \sum_{ij}(2M_j N_i \rho N_i M_j \\
+ 2N_i M_j \rho M_j N_i + [M_i, N_i]\rho[M_j, N_i]) \\
+ \frac{|-\rangle\langle-|}{2} \otimes \sum_{ij}(-[M_j, N_i]\rho[M_j, N_i]). \quad (2)
\end{aligned}
$$

$\rho_c$ is then measured in the Fourier ($\{|+\rangle, |-\rangle\}$) basis and the state correlated with the $|+\rangle$ measurement outcome is post-selected. The benefit of having a quantum switch is due to nontrivial commutations of the set of Kraus operators $\{M_j\}$ and $\{N_i\}$ ($[M_j, N_i] \neq 0$ for some $i, j$). To see why this is the case, suppose the opposite is true, i.e. $[M_j, N_i] = 0$ for all $i, j$, then (2) is equal to $|+\rangle\langle+| \otimes \sum_{ij} M_j N_i \rho N_i^\dagger M_j^\dagger$, which is simply the case of $\rho$ passing through two channels in a definite causal order.

## III. ENTANGLEMENT DISTILLATION PROTOCOLS

Quantum entanglement is a useful resource that is extensively utilized in many quantum technologies, such as quantum clock synchronization [25], device-independent quantum key distribution [26,27], quantum metrology [28], and distributed quantum computing [29]. In the above application settings, constituent particles of the entanglement are usually generated via local physical processes which result in quantum correlations between degrees of freedom of the constituent particles. The particles are then shared between spatially separated parties by being sent through communication channels. In practice, the communication channels are often noisy which degrades the quality of entanglement. In this paper, we focus on the case where entanglement occurs between qubits. A common noise model results in the shared faulty pairs (which we denote as $\chi_i$) being mixtures of Bell states [30]:

$$
\begin{aligned}
\chi_i = A_i|\Phi^+\rangle\langle\Phi^+| + B_i|\Psi^-\rangle\langle\Psi^-| \\
+ C_i|\Psi^+\rangle\langle\Psi^+| + D_i|\Phi^-\rangle\langle\Phi^-|, \quad (3)
\end{aligned}
$$

where

$$|\Phi^{\pm}\rangle = \frac{1}{\sqrt{2}}(|00\rangle \pm |11\rangle), \tag{4}$$

$$|\Psi^{\pm}\rangle = \frac{1}{\sqrt{2}}(|01\rangle \pm |10\rangle). \tag{5}$$

Alternatively, it is compactly expressed as a column vector: $\chi_i = (A_i, B_i, C_i, D_i)^{\mathsf{T}}$ where T denotes the vector transpose. We define the fidelity of $\chi_i$ as

$$F = \max_{|B\rangle \in \{|\Phi^{\pm}\rangle, |\Psi^{\pm}\rangle\}} \langle B|\chi_i|B\rangle. \tag{6}$$

When $\chi_i$ is given in Eq. (3), $F(\chi_i) = \max\{A_i, B_i, C_i, D_i\}$.

Entanglement distillation is a basic protocol that seeks to improve the quality of faulty entangled pairs distributed via noisy processes. In this paper, we focus on two-way entanglement distillation protocols [22]. This type of protocols involves local unitary operations, local measurements, two-way classical communication of measurement results between the communication parties and postselections. It is known that there is a correspondence between two-way entanglement distillation protocols and stabilizer quantum error detecting codes [31]. As a result, the former are usually constructed from the latter. A stabilizer error-detection protocol involves the sender encoding a logical state into a larger Hilbert space using extra ancillas before sending all physical states to the receiver via a noisy channel. The receiver decodes the logical state by measuring the stabilizer of the code. The decoded logical state is kept if the obtained error syndrome signals no error on the decoded state, or discarded if otherwise. In the corresponding two-way entanglement distillation protocol, multiple entangled pairs are shared between the two parties. The receiver performs the decoding circuit of the error-detecting code and the sender performs the complex conjugate of the decoding circuit. Syndrome measurements of the code are then performed by both parties. If the pre-shared entanglements are perfect $|\Phi^+\rangle = 1/\sqrt{2}(|00\rangle + |11\rangle)$, then the syndrome measurements on both parties are perfectly correlated. Errors on the shared entanglement result in a finite amount of $|\Psi^-\rangle$, $|\Psi^+\rangle$ and $|\Phi^-\rangle$ as mixture components, which cause imperfect correlations. Nevertheless, with the presence of errors there is still a finite probability of obtaining the same syndromes on both sides, which projects the unmeasured state into a less-erroneous subspace, or equivalently an entangled state of higher fidelity. We now give a review of various two-way entanglement distillation protocols constructed from small error-detecting codes. These protocols distill a small number of faulty entangled states into one entangled state with a higher fidelity.

## A. DEJMPS protocol

The DEJMPS protocol [21] is an early proposed entanglement distillation protocol that turns two entangled pairs into one. We denote the density matrices of the two entangled pairs as $\chi_0$ and $\chi_1$. Their Hilbert spaces are denoted as $\mathcal{H}_0 = \mathcal{H}_{0A} \otimes \mathcal{H}_{0B}$, where Hilbert space $\mathcal{H}_{0A}$ contains the state of the particle of $\chi_0$ held by one party called Alice and $\mathcal{H}_{0B}$ contains that held by the other party called Bob. Likewise, we define

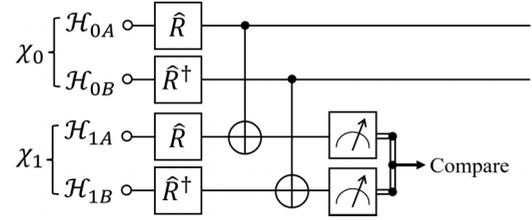

FIG. 1. The circuit for the DEJMPS entanglement distillation protocol. $\chi_1$, $\chi_2$ denote density matrices of the two shared entangled pairs. $\chi_i$ has its two particles sitting in Hilbert spaces $\mathcal{H}_{iA}$ held by Alice and $\mathcal{H}_{iB}$ held by Bob.

$\mathcal{H}_1 = \mathcal{H}_{1A} \otimes \mathcal{H}_{1B}$ for the two particles of $\chi_1$. The protocol has a circuit shown in Fig. 1 and is described as follows:

(1) Alice performs a rotation $\hat{R} = \exp(-i\frac{\pi}{2}\hat{X})$ on the particles on her side. Bob performs the rotation $\hat{R}^{\dagger} = \exp(+i\frac{\pi}{2}\hat{X})$ on the particles on his side.

(2) Alice and Bob each performs a CNOT between the two particles on their sides respectively. The two CNOTs are controlled by (and also targetting) qubits in the same entangled pair,

(3) Alice and Bob measure the target qubits of their CNOTs in the computational basis. Alice then sends Bob her measurement result. If their results agree, the entangled state serving as control qubits in the previous CNOTs is kept. If their results differ, the control pair is discarded and they start over again. In either case, the target pair is also discarded.

The protocol can be described as a CP trace-decreasing map which can be expressed as

$$\mathcal{D}(\chi_0 \otimes \chi_1) = \sum_i \hat{O}^i(\chi_0 \otimes \chi_1)\hat{O}^{i\dagger}, \tag{7}$$

where Kraus operators

$$\hat{O}^i = \langle i|_{1A,1B} \text{CNOT}_{0A}^{1A} \text{CNOT}_{0B}^{1B} \hat{R}_{0A,1A} \hat{R}_{0B,1B}^{\dagger} \tag{8}$$

with $i \in \{00, 11\}$. $\text{CNOT}_{0A}^{1A}$ is a CNOT gate with a control qubit in $\mathcal{H}_{0A}$ and target qubit in $\mathcal{H}_{1A}$ and $\hat{R}_{0A,1A} = \hat{R}_{0A} \otimes \hat{R}_{1A}$ where subscripts denote the Hilbert space of the operators. Equation (7) maps the combined input state in $\mathcal{H}_{0A} \otimes \mathcal{H}_{0B} \otimes \mathcal{H}_{1A} \otimes \mathcal{H}_{1B}$ onto the (unnormalized) output state in $\mathcal{H}_{0A} \otimes \mathcal{H}_{0B}$. The summation is a mixture over the two possible measurement outcomes 00 and 11 that show even parity.

Under the assumption of $\chi_0$ and $\chi_1$ both being Bell-diagonal states, Eq. (7) can also be expressed as

$$\mathcal{D}'(\chi_0 \otimes \chi_1) = \hat{O}(\chi_0 \otimes \chi_1)\hat{O}^{\dagger} \tag{9}$$

where $\hat{O} = \sqrt{2}\hat{O}^{00}$. This simplification is justified by the fact that the measurement projecting the state in $\mathcal{H}_{1A} \otimes \mathcal{H}_{1B}$ onto $|11\rangle$ introduces an extra "$-1$" global phase onto the state in the leftover Hilbert space, and that the extra phases produced on both the "bra" and "ket" side cancel out. This makes projecting onto $|11\rangle$ equivalent to projecting onto $|00\rangle$.

In practice, the various applications which entangled pairs are subject to often request high-fidelity, which can be obtained from repeated applications of single DEJMPS steps using $x > 2$ faulty pairs. This can alternatively be viewed as a single faulty pair undergoing multiple steps of distillation, where each step can be defined with respect to the entangled pair that the original one is combined with. We may then define the causal order of distillation to be the order with which the original pair is combined with others. As an example, for three faulty entangled states $\chi_i$ (where $i \in \{1, 2, 3\}$), we may distill $\chi_3$ and $\chi_2$ into a product of higher fidelity which is then distilled with $\chi_1$. The second distillation step is clearly in the causal future of the first step, as it requires the product of the first step as its input. Likewise, the first step is in the causal past of the second step.

An $n$-to-one distillation scheme when $n > 3$ has more possible arrangements of single DEJMPS steps compared with the case when $n = 3$. As an example, below are all the arrangements when $n = 4$.

(1) Select the supplied pair with the maximum fidelity and discard other three. No distillation is done.

(2) Perform one distillation step using two of the four pairs and discard the other two pairs.

(3) Select three pairs and discard the fourth pair. Within the three chosen ones, select two to perform one step of distillation, the product of which is teamed up with the third pair for another step.

(4) Group all four pairs into teams of two, where one step of distillation is carried out separately for each team before the two products are teamed up for another step. This is called a "recurrence-like structure" [32].

(5) Select two pairs to do one step of distillation, the product of which is purified by combining it with either of the two unchosen pairs before the product is combined with the last unchosen pair for a third step. This is called a "pumpinglike structure" [32,33].

For four faulty pairs $\chi_i$ (where $i \in \{0, 1, 2, 3\}$), we denote the set which contains all the above distillation arrangements as $\mathcal{G}$.

In this paper, we are concerned with the use of entangled states in near-term quantum communication tasks. The most important metrics of two-way entanglement distillation protocols are the fidelities of the output states and their probabilities of success. Apart from this, we recognize another important metric being the number of memories required to carry out the protocol. Quantum states involved in various quantum communication tasks will likely be stored inside memory-based quantum repeaters. Although it is deemed that in the long term, quantum repeaters will be based on error-correcting codes which transmit quantum states fault-tolerantly, in the near term they have fewer memories, high memory noise, infidelity of quantum gates, and high photon loss which are unable to be corrected via error-correction. Instead, they are expected to rely on heralded entanglement generation and two-way entanglement distillation to reduce errors [34,35]. Having an entanglement distillation protocol that uses fewer quantum memories leaves more vacant spaces to receive incoming generated entangled pairs, hence boosts the rate of entanglement distribution.

We note that the concatenated DEJMPS protocols in set $\mathcal{G}$ all use four entangled pairs, but also only require maximally three memory units (for each communication party) to perform. Consider the "recurrence" structure, where the entangled pair distilled from the first DEJMPS step and two other newly generated pairs taking part in the parallel DEJMPS step are stored. Although a total of four entangled pairs are used in this arrangement, only a maximum of three pairs need to be stored in the memories at any time. Later, in Sec. V, we will introduce a class of protocols that simulates two DEJMPS steps applied in an indefinite causal order. Such protocols also make use of four entangled pairs, but only require three entangled pairs to be stored at any time. Nevertheless, there do exists other four-entangled-state protocols that satisfy this storage requirement. We describe them as follows.

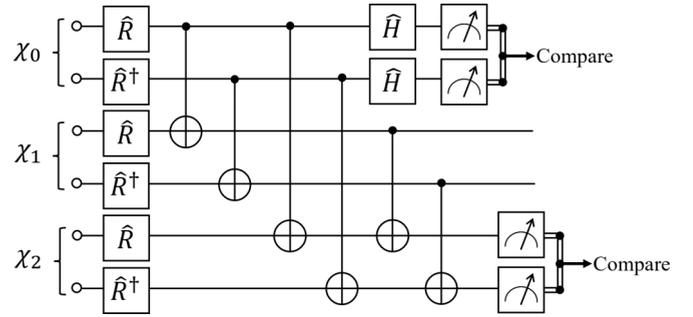

FIG. 2. A three-pair distillation protocol constructed from a three-bit error detecting code stabilized by $\hat{S}_1 = \hat{\mathbb{1}}\hat{Z}\hat{Z}$ and $\hat{S}_2 = \hat{X}\hat{X}\hat{X}$, with an extra initial rotation $\hat{R} = \exp(-i\frac{\pi}{4}\hat{X})$.

### B. Three-pair distillation protocols

The DEJMPS protocol described as above is constructed from the two-bit repetition code (up to an initial single-qubit rotation $\hat{R}$), whose codeword is stabilized by $\hat{Z}\hat{Z}$, where $\hat{Z}$ is the Pauli $Z$ matrix. As a result, the decoding circuit of the code (also up to the rotation $\hat{R}$) is performed on both sides of the entangled pairs to detect errors that do not commute with the stabilizer. When >2 pairs are shared, larger entanglement distillation schemes can be constructed from larger error-detecting codes. We consider a three-qubit code which is stabilized by $\hat{S}_1 = \hat{\mathbb{1}}\hat{Z}\hat{Z}$ and $\hat{S}_2 = \hat{X}\hat{X}\hat{X}$. Such a code has the following decoding circuit

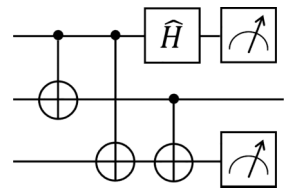

where the top measurement obtains the syndrome $\hat{S}_2$ and the bottom measurement obtains the syndrome $\hat{S}_1$. This leads to the following three-pair entanglement distillation circuit shown in Fig. 2 for three entangled pairs $\chi_0$, $\chi_1$, and $\chi_2$. We have added the same extra rotation $\hat{R} = \exp(-i\frac{\pi}{4})$ to be consistent with the DEJMPS protocol. When four pairs are shared, there are also the following multiple possible arrangements to distill them into one pair. We denote the set which contains all the following arrangements as $\mathcal{J}$:

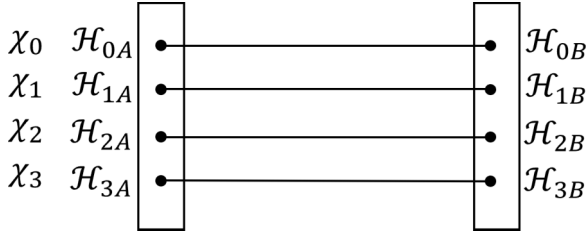

$\chi_0$    $\mathcal{H}_{0A}$      $\mathcal{H}_{0B}$

$\chi_1$    $\mathcal{H}_{1A}$      $\mathcal{H}_{1B}$

$\chi_2$    $\mathcal{H}_{2A}$      $\mathcal{H}_{2B}$

$\chi_3$    $\mathcal{H}_{3A}$      $\mathcal{H}_{3B}$

FIG. 3. Four faulty Bell-diagonal pairs are shared between two repeater stations. $\mathcal{H}_{x,Y}$ where $x \in \{0, 1, 2, 3\}$ and $Y \in \{A, B\}$ denotes the Hilbert space of each stored qubit.

(1) Select three pairs, arrange them in some permutation and distill them into one pair using Fig. 2's circuit. Discard the fourth pair.

(2) Select three pairs and distill them into one using Fig. 2's circuit. A further DEJMPS step is performed on the distilled product and the fourth pair.

(3) Select two pairs, distill them into one pair using DE-JMPS protocol. The product and the two remaining pairs are distilled into one using Fig. 2's circuit.

## IV. APPLYING TWO DEJMPS STEPS IN A SUPERPOSITION OF CAUSAL ORDERS

All protocols $\mathcal{P} \in \mathcal{G}$ and $\mathcal{P} \in \mathcal{J}$ are examples of distillation steps carried out in a definite causal order, where the causal orders between different steps are well defined. In this section, we introduce our modification to break this causality.

### A. Circuit that coherently controls the order of two DEJMPS steps

We consider the following process: the four Bell-diagonal faulty pairs $\chi_i$ are shared between Alice and Bob as illustrated in Fig. 3. Under some control system being in a logical state $|\mathbf{0}\rangle$, $\chi_3$ undergoes a DEJMPS step with $\chi_2$ first, whose dis-tillation output then undergoes another DEJMPS step with $\chi_1$. When the control system is in state $|\mathbf{1}\rangle$ orthonormal to $|\mathbf{0}\rangle$, $\chi_3$ is routed in the quantum registers to first combine with $\chi_1$, whose distillation output is then combined with $\chi_2$. We let the logical states $|\mathbf{0}\rangle$ ($|\mathbf{1}\rangle$) be encoded in the entangled pair $\chi_0$ as $|00\rangle$ ($|11\rangle$) and refer to $\chi_0$ as the "control pair" of the protocol. The above process then can be realized by part of the circuit shown in Fig. 4 before the vertical dashed line. In Fig. 4, each entangled pair $\chi_i$ is inside Hilbert space $\mathcal{H}_{iA} \otimes \mathcal{H}_{iB}$. We denote $\rho_{\text{in}} = \chi_1 \otimes \chi_2 \otimes \chi_3$. One can show before the vertical dashed line, the overall state in Hilbert space $\mathcal{H}_{0A} \otimes \mathcal{H}_{0B} \otimes \mathcal{H}_{1A} \otimes \mathcal{H}_{1B}$, which consists of the control pair and the bipartite state distilled from the two DEJMPS steps, can be expressed as

$$\rho_s = \sum_{i,j=00}^{11} \hat{E}_{ij}(\chi_0 \otimes \rho_{\text{in}})\hat{E}_{ij}^\dagger + \sum_{i,j=00}^{11} \hat{K}_{ij}(\chi_0 \otimes \rho_{\text{in}})\hat{K}_{ij}^\dagger, \quad (10)$$

where

$$\hat{E}_{ij} = |00\rangle\langle00| \otimes \hat{\mathcal{E}}_j^{(2)}\hat{\mathcal{E}}_i^{(1)} + |11\rangle\langle11| \otimes \hat{\mathcal{E}}_i^{(1)}\hat{\mathcal{E}}_j^{(2)}, \quad (11)$$

$$\hat{K}_{ij} = |01\rangle\langle01| \otimes \hat{\mathcal{K}}_i^{(1)} + |10\rangle\langle10| \otimes \hat{\mathcal{K}}_j^{(2)} \quad (12)$$

with

$$\hat{\mathcal{E}}_i^{(1)} = \text{SWAP}_{3A}^{2A}\text{SWAP}_{3B}^{2B}\langle i|_{3A,3B}\text{CNOT}_{2A}^{3A}\text{CNOT}_{2B}^{3B}$$
$$\times \hat{R}_{2A,3A}\hat{R}_{2B,3B}^\dagger, \quad (13)$$

$$\hat{\mathcal{E}}_j^{(2)} = \text{SWAP}_{3A}^{1A}\text{SWAP}_{3B}^{1B}\langle j|_{3A,3B}\text{CNOT}_{1A}^{3A}\text{CNOT}_{1B}^{3B}$$
$$\times \hat{R}_{1A,3A}\hat{R}_{1B,3B}^\dagger, \quad (14)$$

where $\text{SWAP}_{3A}^{2A}$ is a SWAP gate between the states in Hilbert spaces $\mathcal{H}_{3A}$ and $\mathcal{H}_{2A}$. $\hat{\mathcal{K}}_i^{(1)}$ and $\hat{\mathcal{K}}_j^{(2)}$ are the operations on $\rho_{\text{in}}$ in the case that the control pair $\chi_0$ is in the state $|01\rangle$ or $|10\rangle$. We do not give the explicit expression of $\hat{\mathcal{K}}_i^{(1)}$ and $\hat{\mathcal{K}}_j^{(2)}$ as they do not contribute to the discussion. $\hat{\mathcal{E}}_i^{(m)}$ is essentially the Kraus operator of a DEJMPS step as given in Eq. (7) but with extra SWAPs that do not affect outcome of distillation. $\hat{E}_{ij}$ has the form of different basis states of the control pair correlated with opposite orders of $\hat{\mathcal{E}}_i^{(1)}$ and $\hat{\mathcal{E}}_j^{(2)}$, the Kraus operators of the two DEJMPS steps. This clearly shows that the two distillation steps are put into an indefinite causal order. The second term in Eq. (10) occurs when the control pair $\chi_0$ has Bell components $|\Psi^+\rangle\langle\Psi^+|$ and/or $|\Psi^-\rangle\langle\Psi^-|$. This gives rise to another completely positive map which cannot be easily interpreted as $\chi_3$ going through two DEJMPS steps in any order.

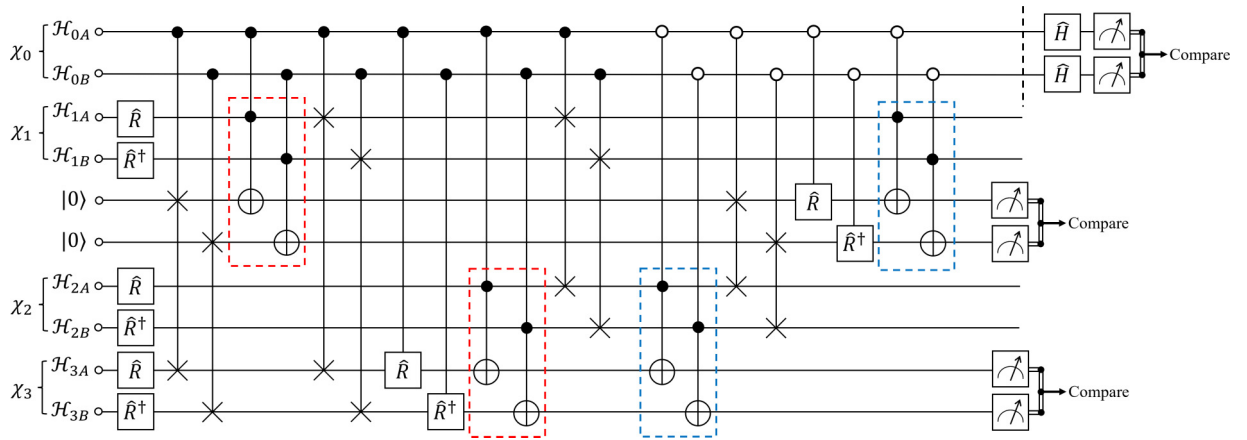

FIG. 4. A modified entanglement distillation scheme which turns four faulty Bell-diagonal states $\chi_i$ (each residing in Hilbert space $\mathcal{H}_{iA} \otimes \mathcal{H}_{iB}$) into one. The circuit features two DEJMPS distillation steps applied in a superposition of causal orders which is coherently controlled by the faulty pair $\chi_0$. The red and blue dashed rectangles encircle the CNOT gates of the two DEJMPS steps applied in the $\chi_3 \to \chi_1 \to \chi_2$ ($\chi_3 \to \chi_2 \to \chi_1$) causal order. After the vertical dashed line in the figure, we postselect the distilled state upon receiving an even parity measurement outcome from the state in $\mathcal{H}_{0A} \otimes \mathcal{H}_{0B}$. This interferes the products distilled from the two causal orders with the hope of boosting the fidelity of the final product of distillation.

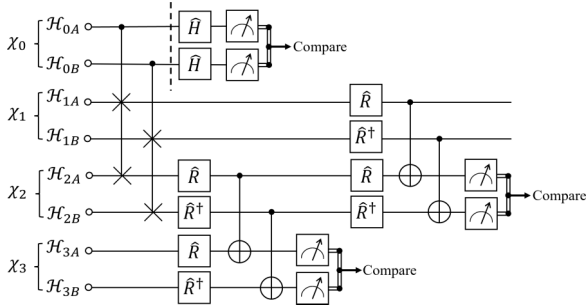

FIG. 5. A simple circuit which seeks to simulate two DEJMPS steps in a coherent superposition of two causal orders. This is achieved by coherent swapping of $\chi_1$ and $\chi_2$ at the beginning of the circuit, in contrast to Fig. 4 where $\chi_1$ and $\chi_2$ are kept still and $\chi_3$ is routed around. $\hat{R} = \exp(-i\frac{\pi}{2}\hat{X})$ where $\hat{X}$ is the Pauli-X operator. $\hat{R}^\dagger$ is the Hermitian conjugate of $\hat{R}$. $\hat{H}$ is the Hadamard gate.

After the two DEJMPS steps, a Hadamard gate is applied onto both particles of the control pair, which is followed by a parity measurement on the control pair (shown in Fig. 4 after the vertical dashed line). We then postselect the bipartite state in the unmeasured Hilbert space $\mathcal{H}_{1A} \otimes \mathcal{H}_{1B}$ over an even parity outcome. This allows the distillation product from the two causal orders to interfere and the state in the unmeasured Hilbert space is the final product of the protocol.

One sees from Fig. 4 that the circuit is quite complicated. It also requires the control pair to coherently control every gate within the DEJMPS protocol. This gives rise to a large number of double-controlled gates between three qubits, which are difficult to implement. These factors make the circuit difficult to demonstrate as an entanglement distillation protocol. In Sec. V, we introduce a much simpler circuit that also simulates two DEJMPS steps in an indefinite causal order. In this new circuit, the causal orders of DEJMPS steps are not emulated by routing $\chi_3$, but merely control-SWAPPING $\chi_1$ and $\chi_2$. We show that the output state under this simplified protocol can be expressed as $\rho_{in}$ undergoing two maps with set of Kraus operators defined in Eq. (9). The new circuit uses a smaller number of gates, especially many fewer controlled gates among three particles. These features shall make the new circuit easier to demonstrate in practice.

## V. SIMPLIFIED COHERENT CONTROL OF ORDER OF TWO DEJMPS STEPS

The simplified scheme of coherently controlling the order of two DEJMPS distillation steps has a circuit shown in Fig. 5 and is described as follows.

(1) We apply a controlled-SWAP gate from the particle in Hilbert space $\mathcal{H}_{0A}$ to the two particles in Hilbert spaces $\mathcal{H}_{1A}$ and $\mathcal{H}_{2A}$. The particle in Hilbert space $\mathcal{H}_{0B}$ performs a controlled-SWAP gate on the two particles in Hilbert spaces $\mathcal{H}_{1B}$ and $\mathcal{H}_{2B}$. The controlled-SWAP gate is such that the two target states are swapped if the control qubit is in $|1\rangle$.

(2) We apply one step of DEJMPS protocol on Hilbert spaces $\mathcal{H}_{3A} \otimes \mathcal{H}_{3B}$ and $\mathcal{H}_{2A} \otimes \mathcal{H}_{2B}$. If the measurement outcomes show even parity, $\mathcal{H}_{2A}$ and $\mathcal{H}_{2B}$ are kept and we proceed to step 3. Otherwise we discard all pairs and start the scheme over with newly supplied faulty pairs.

(3) We apply another step of Deutsch's protocol on Hilbert spaces $\mathcal{H}_{2A} \otimes \mathcal{H}_{2B}$ and $\mathcal{H}_{1A} \otimes \mathcal{H}_{1B}$. If the parity measurement outcome is even, $\mathcal{H}_{1A}$ and $\mathcal{H}_{1B}$ are kept and we continue to step 4. Otherwise we discard all pairs and start again from the beginning.

(4) We measure $\mathcal{H}_{0A}$ and $\mathcal{H}_{0B}$ separately in the Fourier $\{|+\rangle, |-\rangle\}$ basis and compare the results. If they show even parity, the scheme is successful. Otherwise we discard all pairs and start again from the beginning.

Although we have described operations during the protocol as occurring among certain Hilbert spaces, the above protocol also works for other combinations of the Hilbert spaces. For example, the control-SWAP in step 1 can instead be controlled by particles in Hilbert spaces $\mathcal{H}_{1A} \otimes \mathcal{H}_{1B}$ which targets $\mathcal{H}_{2A} \otimes \mathcal{H}_{2B}$ and $\mathcal{H}_{3A} \otimes \mathcal{H}_{3B}$. Then the two DEJMPS steps are carried out between states in $\mathcal{H}_{0A} \otimes \mathcal{H}_{0B}$ & $\mathcal{H}_{2A} \otimes \mathcal{H}_{2B}$ and $\mathcal{H}_{2A} \otimes \mathcal{H}_{2B}$ & $\mathcal{H}_{3A} \otimes \mathcal{H}_{3B}$. This variation defines a whole set of protocols (there are in fact 12 such protocols), which we denote as $\mathcal{S}$, each having a similar structure.

### A. Difference between routing $\chi_3$ and swapping $\chi_1$ and $\chi_2$

One may be tempted to infer that the two circuits in Figs. 4 and 5 are equivalent: whether routing $\chi_3$ around while keeping $\chi_2$ and $\chi_1$ still (in Fig. 4) or swapping $\chi_2$ and $\chi_1$ around while keeping $\chi_3$ still (in Fig. 5) should not matter, as they both simulate $\chi_3$ going through DEJMPS distillations with $\chi_2$ and $\chi_1$ in two opposite orders. This is not actually the case, due to the parity measurements that are part of the distillation protocol. In Fig. 4's circuit, the two DEJMPS steps, each associated with $\chi_2$ or $\chi_1$, can have different parity measurement results (00 or 11, under the even-parity requirement), and that the parity measurement results associated with $\chi_2$ and $\chi_1$ are the same for both causal orders (whether $\chi_2$ is used after or before $\chi_1$). In Fig. 5's circuit, however, the first DEJMPS steps (regardless of whether it is with $\chi_2$ or $\chi_1$) performed in both causal branches have the same parity measurement outcome, and so do the second DEJMPS steps in both causal branches. This means that if the CP trace-decreasing map of each DEJMPS step is expressed as Eq. (7), which is a sum over projections onto $|00\rangle$ and $|11\rangle$, the overall state before the vertical dashed line in Fig. 5's circuit cannot be written as Eq. (10), where the effective Kraus operator $\hat{E}_{ij}$ has components that are exact swapping of the Kraus operators of the two DEJMPS steps. We therefore cannot interpret the distillation output of Fig. 5's circuit as $\chi_3$ going through two DEJMPS steps in an indefinite causal order, if the completely positive map of the DEJMPS protocol is given as Eq. (7).

### B. Completely positive trace-decreasing maps under swapping of $\chi_1$ and $\chi_2$

In this section, we are able to show that Fig. 5's circuit can be interpreted as $\chi_3$ undergoing two DEJMPS steps in an indefinite causal order, if the DEJMPS protocol is instead described as being described by Eq. (9). We point out that for Bell-diagonal states as being considered in this paper, both (7) and (9) are equally valid quantum maps describing the DEJMPS protocol since mathematically they produce the exact distillation output as defined by the DEJMPS circuit [21]. In this section, we assume the control pair is $\chi_0$ which control-SWAPs $\chi_1$ and $\chi_2$. This makes $\rho_0 = \chi_0 \otimes \chi_1 \otimes \chi_2 \otimes \chi_3$ and $\rho_{in} = \chi_1 \otimes \chi_2 \otimes \chi_3$.

Following the circuit in Fig. 5, one can show that at the dashed line before the final Hadamard gates and parity measurements, the state in $\mathcal{H}_{0A} \otimes \mathcal{H}_{0B} \otimes \mathcal{H}_{1A} \otimes \mathcal{H}_{1B}$ reads

$$\rho_s = \frac{A_0 + D_0}{2}(|00\rangle\langle 00| \otimes N_1 + |11\rangle\langle 11| \otimes N_2)$$
$$+ \frac{A_0 - D_0}{2}(|00\rangle\langle 11| \otimes M_1 + |11\rangle\langle 00| \otimes M_2)$$
$$+ \frac{B_0 + C_0}{2}(|01\rangle\langle 01| \otimes T_1 + |10\rangle\langle 10| \otimes T_2)$$
$$+ \frac{C_0 - B_0}{2}(|01\rangle\langle 10| \otimes L_1 + |10\rangle\langle 01| \otimes L_2), \quad (15)$$

where

$$N_1 = \sum_{i,j \in 00,11} \hat{F}^j \hat{O}^i \rho_{in} \hat{O}^{i\dagger} \hat{F}^{j\dagger}, \quad (16)$$

$$N_2 = \sum_{i,j \in 00,11} \hat{F}^j \hat{P}^i \rho_{in} \hat{P}^{i\dagger} \hat{F}^{j\dagger}, \quad (17)$$

$$M_1 = \sum_{i,j \in 00,11} \hat{F}^j \hat{P}^i \rho_{in} \hat{O}^{i\dagger} \hat{F}^{j\dagger}, \quad (18)$$

$$M_2 = \sum_{i,j \in 00,11} \hat{F}^j \hat{O}^i \rho_{in} \hat{P}^{i\dagger} \hat{F}^{j\dagger}, \quad (19)$$

and

$$\hat{O}^i = \langle i|_{3A,3B} \text{CNOT}^{3A}_{2A} \text{CNOT}^{3B}_{2B} \hat{R}_{2A,3A} \hat{R}^\dagger_{2B,3B}, \quad (20)$$

$$\hat{P}^i = \hat{O}^i \text{SWAP}^{2A}_{1A} \text{SWAP}^{2B}_{1B}, \quad (21)$$

$$\hat{F}^j = \langle j|_{2A,2B} \text{CNOT}^{2A}_{1A} \text{CNOT}^{2B}_{1B} \hat{R}_{1A,2A} \hat{R}^\dagger_{1B,2B}. \quad (22)$$

The explicit expressions of $T_1$, $T_2$, $L_1$, and $L_2$ are given in Sec. V C, which can be found by tracing the circuit in Fig. 5 in case the control pair is in state $|01\rangle$ or $|10\rangle$.

*Theorem 1.* $N_1$ and $N_2$, the (unnormalized) distillation product states from two DEJMPS steps in opposite causal orders can be expressed as

$$N_1 = \hat{Q}_2 \hat{Q}_1 \rho_{in} \hat{Q}_1 \hat{Q}_2, \quad (23)$$

$$N_2 = \hat{Q}_1 \hat{Q}_2 \rho_{in} \hat{Q}_2 \hat{Q}_1, \quad (24)$$

while $M_1$ and $M_2$, the (unnormalized) entangled states correlated with the off-diagonal terms between the two basis states of the control pair that C-SWAPs $\chi_1$ and $\chi_2$, can be

expressed as

$$M_1 = \hat{Q}_1 \hat{Q}_2 \rho_{in} \hat{Q}_1 \hat{Q}_2, \quad (25)$$

$$M_2 = \hat{Q}_2 \hat{Q}_1 \rho_{in} \hat{Q}_2 \hat{Q}_1, \quad (26)$$

where Kraus operators

$$\hat{Q}_1 = \sqrt{2}\langle 00|_{2A,2B} \text{SWAP}^{3A}_{2A} \text{SWAP}^{3B}_{2B} \quad \text{CNOT}^{3A}_{2A} \text{CNOT}^{3B}_{2B}$$
$$\times \hat{R}_{2A,3A} \hat{R}^\dagger_{2B,3B}, \quad (27)$$

$$\hat{Q}_2 = \sqrt{2}\langle 00|_{1A,1B} \text{SWAP}^{3A}_{1A} \text{SWAP}^{3B}_{1B} \text{CNOT}^{3A}_{1A} \text{CNOT}^{3B}_{1B}$$
$$\times \hat{R}_{1A,3A} \hat{R}^\dagger_{1B,3B} \quad (28)$$

are equivalent to the Kraus operator of a single DEJMPS step that only projects the parity-measured states onto $|00\rangle$, up to extra SWAP gates that relabel the Hilbert spaces of the subsystems.

*Proof* To show that $N_1$, $N_2$, $M_1$ and $M_2$ can be expressed as Eqs. (23) to (26), two major steps are carried out. We first show that both outcomes of parity measurements (00 and 11) yield the same distillation output. This allows us to express mixture over projections onto "00" and "11" as only projecting onto "00." Secondly, we intersperse gates inside $\hat{O}^i$, $\hat{P}^i$, and $\hat{F}^i$ with additional SWAP gates to relabel some of the involved Hilbert spaces in order to arrive at $\hat{Q}_1$ and $\hat{Q}_2$.

For convenience of description, we introduce a more compact notation that denotes $\chi_i$ as

$$\chi_i = \sum_{a,b} m^{(i)}_{a,b} |\beta_{a,b}\rangle\langle\beta_{a,b}|, \quad (29)$$

where $a \in \{0, 1\}$ denotes the parity of the Bell component and $b \in \{0, 1\}$ denotes the "+," "−," sign of the component such that $\{|\beta_{0,0}\rangle, |\beta_{1,1}\rangle, |\beta_{1,0}\rangle, |\beta_{0,1}\rangle\} = \{|\Phi^+\rangle, |\Psi^-\rangle, |\Psi^+\rangle, |\Phi^-\rangle\}$. $m^{(i)}_{a,b}$s are the corresponding coefficients of the components which satisfy $\sum_{a,b} m^{(i)}_{a,b} = 1$ due to normalization. The input faulty states $\rho_{in}$ can be expressed as a mixture over components, each being a tensor product of three Bell states:

$$\rho_{in} = \sum_{\mathbf{a},\mathbf{b}} \left(\prod_i m^{(i)}_{a_i,b_i}\right) \Gamma_{\mathbf{a},\mathbf{b}}, \quad (30)$$

where $\Gamma_{\mathbf{a},\mathbf{b}} = |\beta_{a_1,b_1}\rangle|\beta_{a_2,b_2}\rangle|\beta_{a_3,b_3}\rangle \times$ h.c and $\mathbf{a} = (a_1, a_2, a_3)$, $\mathbf{b} = (b_1, b_2, b_3)$.

It is not immediately clear that the two parity measurement outcomes (00 and 11) are correlated with equivalent output state and as a result, it is not clear that the sum over "00" and "11" in Eqs. (16)–(19) can just be re-expressed as projecting only onto "00." To see this, we first examine the inner parts of $N_1$, $N_2$ and $M_1$: $\hat{O}^i \Gamma_{\mathbf{a},\mathbf{b}} \hat{O}^{i\dagger}$, $\hat{P}^i \Gamma_{\mathbf{a},\mathbf{b}} \hat{P}^{i\dagger}$ and $\hat{P}^i \Gamma_{\mathbf{a},\mathbf{b}} \hat{O}^{i\dagger}$ ($M_2$ is simply the Hermitian conjugate of $M_1$). These terms consist of initial rotations $\hat{R}$ on two Bell states, CNOTs and projection onto $|00\rangle$ or $|11\rangle$. It is known from Ref. [21] that $\hat{R}$ preserves the diagonal structure of $\Gamma_{\mathbf{a},\mathbf{b}}$: it merely permutes $|\Phi^-\rangle$ and $|\Psi^-\rangle$, leaving $|\Phi^+\rangle$ and $|\Psi^+\rangle$ unchanged. The trailing CNOT gates hence still act on a Bell-diagonal state. The CNOTs turn $|\beta_{a_2,b_2}\rangle|\beta_{a_3,b_3}\rangle$ into $|\beta_{a_2,b_2\oplus b_3}\rangle|\beta_{a_2\oplus a_3,b_3}\rangle$ and preserve the sign of the Bell state ($b_3$) in $\mathcal{H}_{3A} \otimes \mathcal{H}_{3B}$, where the subsequent first parity measurement is done. The sign of the Bell state $b_3$ is important. We note that projections of $|\Phi^+\rangle$ (the case where $b_3 = 0$) onto $|00\rangle$ and $|11\rangle$ both yield a trivial global phase. But for the "negative sign" $|\Phi^-\rangle$ (where $b_3 = 1$), projection onto $|00\rangle$ yields a trivial global phase, while projection onto $|11\rangle$ yields a "$-1$" global phase. Here, since the terms $\hat{O}^i\Gamma_{\mathbf{a},\mathbf{b}}\hat{O}^{i\dagger}$, $\hat{P}^i\Gamma_{\mathbf{a},\mathbf{b}}\hat{P}^{i\dagger}$ and $\hat{P}^i\Gamma_{\mathbf{a},\mathbf{b}}\hat{O}^{i\dagger}$ are correlated with the control pair being in $|00\rangle\langle00|$, $|11\rangle\langle11|$, and $|00\rangle\langle11|$, respectively [see Eqs. (15)–(19)], the global phases generated during the parity measurement become relative phases among the above three terms, which can affect the final distilled state nontrivially. However, the fact that the CNOTs preserve the sign of the target Bell state means after the CNOTs, the Bell states on the "bra" and "ket" sides must have the same sign ($b_3$) regardless of the control Bell state of the CNOTs on the two sides. The global phases induced on the "00" and "11" outcomes have equivalent effect onto the leftover Hilbert space. Mathematically, the following relations are obtained: $\hat{O}^{00}\Gamma_{\mathbf{a},\mathbf{b}}\hat{O}^{00\dagger} = \hat{O}^{11}\Gamma_{\mathbf{a},\mathbf{b}}\hat{O}^{11\dagger}$, $\hat{P}^{00}\Gamma_{\mathbf{a},\mathbf{b}}\hat{P}^{00\dagger} = \hat{P}^{11}\Gamma_{\mathbf{a},\mathbf{b}}\hat{P}^{11\dagger}$, and $\hat{P}^{00}\Gamma_{\mathbf{a},\mathbf{b}}\hat{O}^{00\dagger} = \hat{P}^{11}\Gamma_{\mathbf{a},\mathbf{b}}\hat{O}^{11\dagger}$.

The leftover state in $(\mathcal{H}_{1A} \otimes \mathcal{H}_{1B}) \otimes (\mathcal{H}_{2A} \otimes \mathcal{H}_{2B})$, after tracing out $\mathcal{H}_{3A} \otimes \mathcal{H}_{3B}$, is also a tensor product of two Bell states, and is now subject to another round of local rotations, CNOTs and parity measurement. One can use the same argument as above to show that projecting onto $|00\rangle$ and $|11\rangle$ in the parity measurement yield the same state in the leftover Hilbert space $\mathcal{H}_{1A} \otimes \mathcal{H}_{1B}$. This means $\hat{F}^{00}\hat{\Omega}^i\Gamma_{\mathbf{a},\mathbf{b}}\hat{\Upsilon}^{i\dagger}\hat{F}^{00\dagger} = \hat{F}^{11}\hat{\Omega}^i\Gamma_{\mathbf{a},\mathbf{b}}\hat{\Upsilon}^{i\dagger}\hat{F}^{11\dagger}$ for $i \in \{00, 11\}$, where $\hat{\Omega}$ and $\hat{\Upsilon}$ are either $\hat{O}$ or $\hat{P}$. Since this is true for $\Gamma_{\mathbf{a},\mathbf{b}}$ of arbitrary $\mathbf{a}$, $\mathbf{b}$, it is also true for $\rho_{\mathrm{in}}$, which is a mixture of $\Gamma_{\mathbf{a},\mathbf{b}}$ of different $\mathbf{a}$ and $\mathbf{b}$. We can now define $\hat{F} = \sqrt{2}\hat{F}^{00}$, $\hat{O} = \sqrt{2}\hat{O}^{00}$ and $\hat{P} = \sqrt{2}\hat{P}^{00}$ and express $N_1$, $N_2$, $M_1$ and $M_2$ as $\rho_{\mathrm{in}}$ going through single Kraus operators, rather than mixtures over Kraus operators that project onto different parity-measurement outcomes 00 or 11:

$$N_1 = \hat{F}\hat{O}\rho_{\mathrm{in}}\hat{O}^\dagger\hat{F}^\dagger, \tag{31}$$

$$N_2 = \hat{F}\hat{P}\rho_{\mathrm{in}}\hat{P}^\dagger\hat{F}^\dagger, \tag{32}$$

$$M_1 = \hat{F}\hat{P}\rho_{\mathrm{in}}\hat{O}^\dagger\hat{F}^\dagger, \tag{33}$$

$$M_2 = \hat{F}\hat{O}\rho_{\mathrm{in}}\hat{P}^\dagger\hat{F}^\dagger. \tag{34}$$

We now want to express $N_1$ as $\rho_{\mathrm{in}}$ passing through two CP trace-decreasing maps in one order, and $N_2$ as $\rho_{\mathrm{in}}$ undergoing the same two maps in the opposite order. We have defined a DEJMPS step, which has its distinct CP trace-decreasing map and Kraus operator, solely with respect to the entangled pair ($\chi_1$ or $\chi_2$) that $\chi_3$ is combined with. In Fig. 5's circuit, however, the faulty state subject to the second distillation step (regardless of which entangled pair it is combined with) is in Hilbert space $\mathcal{H}_{2A} \otimes \mathcal{H}_{2B}$, which is different from the Hilbert space of $\chi_3$ ($\mathcal{H}_{3A} \otimes \mathcal{H}_{3B}$) during the first distillation step. Since operations on distinct Hilbert spaces are described by different operators, this prevents the same Kraus operator that describes the 1st distillation step in one causal order from also describing the 2nd distillation step in the other causal order. To resolve this, we add extra SWAP operations to $\hat{F}$, $\hat{O}$, and $\hat{P}$ such that the state produced from each DEJMPS step is always in Hilbert space $\mathcal{H}_{3A} \otimes \mathcal{H}_{3B}$, regardless of whether the DEJMPS step is done as the first or second in the queue. Consider the term $N_1$. We add two SWAP gates $\mathrm{SWAP}_{2A}^{3A}\mathrm{SWAP}_{2B}^{3B}$ between the CNOTs and parity measurement of operator $\hat{O}$ [in Eq. (20)]. In order for the circuit's output to remain unchanged, this must also change the projected Hilbert spaces of parity measurement of $\hat{O}$ from $\mathcal{H}_{3A} \otimes \mathcal{H}_{3B}$ to $\mathcal{H}_{2A} \otimes \mathcal{H}_{2B}$, and as a result the unmeasured state is now in $\mathcal{H}_{3A} \otimes \mathcal{H}_{3B}$. The modified $\hat{O}$ is shown to equal $\hat{Q}_1$. The Hilbert spaces on which gates in the subsequent Kraus operator $\hat{F}$ act are also swapped between $\mathcal{H}_{3A} \otimes \mathcal{H}_{3B}$ and $\mathcal{H}_{2A} \otimes \mathcal{H}_{2B}$. We then add two other SWAP gates $\mathrm{SWAP}_{1A}^{3A}\mathrm{SWAP}_{1B}^{3B}$ between the CNOTs and parity measurement of $\hat{F}$, which changes the parity-measured Hilbert space to $\mathcal{H}_{1A} \otimes \mathcal{H}_{1B}$. This turns $\hat{F}$ into $\hat{Q}_2$. We emphasize that the additional SWAP gates are introduced merely for algebraic reasons and are not implemented physically.

As for the term $N_2$ in Eq. (32), one can see from the construction of $\hat{O}^i$ and $\hat{P}^i$ (hence $\hat{O}$ and $\hat{P}$) in Eqs. (20) and (21) that $N_2$ is essentially $N_1$ with a swapping of labels of Hilbert spaces $\mathcal{H}_{2A} \otimes \mathcal{H}_{2B}$ and $\mathcal{H}_{1A} \otimes \mathcal{H}_{1B}$. One can then conclude that if the same procedure of addition of SWAP gates is carried out on $N_2$, the result will be that on $N_1$ [in Eq. (23)] also followed by a relabelling of Hilbert spaces $\mathcal{H}_{2A} \otimes \mathcal{H}_{2B}$ and $\mathcal{H}_{1A} \otimes \mathcal{H}_{1B}$, which is exactly equal to an exchange of $\hat{Q}_1$ and $\hat{Q}_2$, leading to Eq. (24). The manipulations on $M_1$ and $M_2$ follow a similar description that leads to Eqs. (25) and (26). One can see from Eqs. (27) and (28) that $\hat{Q}_1$ and $\hat{Q}_2$ are essentially the Kraus operator $\hat{O}$ in Eq. (9) but with extra SWAP gates that relabel the Hilbert spaces of some of the states which do not affect the distillation output.

According to Eqs. (23) and (24), one can regard $N_2$ as the input state $\rho_{\mathrm{in}}$ undergoing trace decreasing maps $\hat{Q}_1$ and $\hat{Q}_2$ in the opposite order as that in $N_1$. Overall, the circuit in Fig. 5 before the final Hadamard and parity measurement can be expressed as

$$\rho_s = \hat{W}(\chi_0 \otimes \rho_{\mathrm{in}})\hat{W}^\dagger + \hat{V}(\chi_0 \otimes \rho_{\mathrm{in}})\hat{V}^\dagger, \tag{35}$$

where

$$\hat{W} = |00\rangle\langle00| \otimes \hat{Q}_2\hat{Q}_1 + |11\rangle\langle11| \otimes \hat{Q}_1\hat{Q}_2, \tag{36}$$

$$\hat{V} = |01\rangle\langle01| \otimes \hat{J} + |10\rangle\langle10| \otimes \hat{S}. \tag{37}$$

with $\hat{J}$ and $\hat{S}$ being the operators that act on $\rho_{\mathrm{in}}$ in case the control pair is in state $|01\rangle$ and $|10\rangle$. Equation (36) features opposite orders of $\hat{Q}_1$ and $\hat{Q}_2$ correlated with different states of the control pair. We have hence shown the circuit in Fig. 5 simulates two DEJMPS steps applied in a causal order controlled by $\chi_0$.

### C. Explicit expression of output state from the protocol

After the final two Hadamard gates in Fig. 5, the state in Hilbert space $\mathcal{H}_{0A} \otimes \mathcal{H}_{0B} \otimes \mathcal{H}_{1A} \otimes \mathcal{H}_{1B}$ is expressed as

$$\rho_s = \frac{|00\rangle\langle00| + |11\rangle\langle11|}{4} \otimes \rho_+ + \frac{|01\rangle\langle01| + |10\rangle\langle10|}{4} \otimes \rho_- + \rho_{sr}, \tag{38}$$

where

$$\rho_{\pm} = \frac{A_0 + D_0}{2} N_1 + \frac{A_0 + D_0}{2} N_2 \\ \pm \frac{A_0 - D_0}{2} 2M + \frac{B_0 + C_0}{2} 2T \pm \frac{C_0 - B_0}{2} 2L \quad (39)$$

and $\rho_{sr}$ consists of off-diagonal elements of the state in $\mathcal{H}_{0A} \otimes \mathcal{H}_{0A}$. Upon parity-measuring $\chi_0$, $\rho_{sr}$ vanishes. The unmeasured state is postselected upon an even-parity outcome, making $\rho_+$ the final product of the protocol.

We give explicit expressions of all terms in Eq. (39) as follows:

$$N_1 = \begin{pmatrix} A_1(A_2 A_3 + B_2 B_3) + B_1(D_2 C_3 + C_2 D_3) \\ D_1(C_2 C_3 + D_2 D_3) + C_1(B_2 A_3 + A_2 B_3) \\ C_1(C_2 C_3 + D_2 D_3) + D_1(B_2 A_3 + A_2 B_3) \\ B_1(A_2 A_3 + B_2 B_3) + A_1(D_2 C_3 + C_2 D_3) \end{pmatrix}, \quad (40)$$

$$N_2 = \begin{pmatrix} A_2(A_1 A_3 + B_1 B_3) + B_2(D_1 C_3 + C_1 D_3) \\ D_2(C_1 C_3 + D_1 D_3) + C_2(B_1 A_3 + A_1 B_3) \\ C_2(C_1 C_3 + D_1 D_3) + D_2(B_1 A_3 + A_1 B_3) \\ B_2(A_1 A_3 + B_1 B_3) + A_2(D_1 C_3 + C_1 D_3) \end{pmatrix}, \quad (41)$$

$$M = M_1 = M_2 = \begin{pmatrix} A_3 A_2 A_1 \\ B_3 B_2 B_1 \\ C_3 C_2 C_1 \\ D_3 D_2 D_1 \end{pmatrix}. \quad (42)$$

$T$ and $L$ are expressed as

$$T = \frac{1}{4} \begin{pmatrix} \sum\limits_{a,b,c,d=0}^{1} m_{a,b}^{(1)} m_{c,d}^{(2)} m_{a \oplus c, b \oplus d}^{(3)} \\ \sum\limits_{a,b,c,d=0}^{1} m_{a,b}^{(1)} m_{c,d}^{(2)} m_{a \oplus c \oplus 1, b \oplus d \oplus 1}^{(3)} \\ \sum\limits_{a,b,c,d=0}^{1} m_{a,b}^{(1)} m_{c,d}^{(2)} m_{a \oplus c \oplus 1, b \oplus d}^{(3)} \\ \sum\limits_{a,b,c,d=0}^{1} m_{a,b}^{(1)} m_{c,d}^{(2)} m_{a \oplus c, b \oplus d \oplus 1}^{(3)} \end{pmatrix}, \quad (43)$$

$$L = \frac{1}{4} \begin{pmatrix} \sum\limits_{\substack{a,b,\\c,d=0}}^{1} m_{a,b}^{(1)} m_{c,d}^{(2)} m_{a \oplus c, b \oplus d}^{(3)} (-1)^{a(1 \oplus d) \oplus c(1 \oplus b)} \\ \sum\limits_{\substack{a,b,\\c,d=0}}^{1} m_{a,b}^{(1)} m_{c,d}^{(2)} m_{a \oplus c \oplus 1, b \oplus d \oplus 1}^{(3)} (-1)^{(a \oplus 1)d \oplus (c \oplus 1)b} \\ \sum\limits_{\substack{a,b,\\c,d=0}}^{1} m_{a,b}^{(1)} m_{c,d}^{(2)} m_{a \oplus c \oplus 1, b \oplus d}^{(3)} (-1)^{a(1 \oplus d) \oplus c(1 \oplus b) \oplus b \oplus d} \\ \sum\limits_{\substack{a,b,\\c,d=0}}^{1} m_{a,b}^{(1)} m_{c,d}^{(2)} m_{a \oplus c, b \oplus d \oplus 1}^{(3)} (-1)^{ad \oplus cb} \end{pmatrix}. \quad (44)$$

In Eqs. (43) and (44), the definition of $m_{j,k}^{(i)}$ was given in Eq. (29). $\oplus$ is binary addition acting on $a, b \in \{0, 1\}$ such that $a \oplus b = \text{mod}(a + b, 2)$.

## VI. ADVANTAGE IN FIDELITY AND PROBABILITY OF SUCCESS

We now compare, for four given input faulty states, the output fidelity from our set of protocols $\mathcal{S}$ against the concatenated DEJMPS protocols (denoted as set $\mathcal{G}$) and the three-pair distillation protocols (denoted as set $\mathcal{J}$). We point out that previous works studying indefinite causal structures in quantum information processing tasks have argued their advantage by comparing them against definite causal structures that consist of the same number of elementary CP maps. This can be justified by treating the control state in the quantum switch as free resource, which is reasonable in some experimental setups that are used to implement these tasks (e.g., interferometers, where the control state is simply the propagation paths of photons). Here in our set of protocols $\mathcal{S}$, the control of causal order is carried out by an entangled pair. This extra pair should not be seen as free, but as costly as the other entangled pairs involved in the protocol since they all practically take similar physical resources to generate. Therefore the comparison of protocols $\mathcal{S}$ should not be made against only the concatenated two DEJMPS steps, but against the set of definite-causal-order protocols that also turn four faulty entangled pairs into one. These protocols are exactly those that form the set of protocols $\mathcal{G}$ and $\mathcal{J}$. We note that we do not compare our protocols against those entanglement distillation protocols that utilize four entangled pairs together (such protocols can usually be constructed from four-bit quantum error-detecting codes) as they require four memory units to carry out, which is a higher requirement than the above-mentioned protocols.

For protocols in $\mathcal{S}$, we note that the output of the protocol $\rho_+$ from Eq. (39) is a mixture of five terms: $N_1$, $N_2$, $M$, $T$, and $L$ (where we can express $M$ as $M = \frac{1}{2}(N_1 + N_2 + [\hat{Q}_2, \hat{Q}_1] \rho_{\text{in}} [\hat{Q}_2, \hat{Q}_1])$). In order for $\rho_+$ to have a fidelity advantage over the distillation products of all protocols in $\mathcal{G}$, at least one term among $M$, $T$, and $L$ must have a fidelity larger than both $N_1$ and $N_2$, since $N_1$ and $N_2$ themselves are the product of distillation from two definite-ordered DEJMPS steps, which are member protocols of the set $\mathcal{G}$. In this section, aside from comparing the fidelities and probabilities of success of protocol set $\mathcal{S}$ and $\mathcal{G}$, we present that the advantage in fidelity solely comes from $M$ (rather than from $T$ and $L$), owing to the nontrivial commutation (i.e. $[\hat{Q}_2, \hat{Q}_1] \neq 0$) between the Kraus operators of the two maps that describe the two DEJMPS steps. This is the same origin as that of the advantage of a quantum switch as reviewed earlier in Sec. II. This confirms that the advantage in fidelity of our scheme is indeed due to applying entanglement distillation maps in a coherent superposition of two causal orders.

We first present an discrete example of input state $\chi_i$ where protocol set $\mathcal{S}$ returns a higher fidelity and success probability than protocols in $\mathcal{G}$ and $\mathcal{J}$, showing clear overall advantage of protocols $\mathcal{S}$. It is discovered that the advantage occur on faulty states with noise biases close to that of Werner states. We then present the parameter region of input fidelities where $\mathcal{S}$'s advantage holds assuming the input states are Werner states, and also comment on $\mathcal{S}$'s distillation performance when input states have biased noise.

### A. Discrete advantageous input examples

#### 1. Werner states as input

There is a continuous region of input state parametesr where the advantage holds. We first consider the case when the faulty pairs all experience depolarising noise from $|\Phi^+\rangle$ which turns them into Werner states, which have the form $\chi_i = F_i|\Phi^+\rangle\langle\Phi^+| + (1 - F_i)/3(|\Psi^-\rangle\langle\Psi^-| + |\Psi^+\rangle\langle\Psi^+| + |\Phi^-\rangle\langle\Phi^-|)$. We search over the parameter space $\Sigma = \{\mathbf{F} = [F_0, F_1, F_2, F_3] \mid 0.25 < F_i < 1 \,\forall i\}$. For each set of parameters $\mathbf{F}$, we find, over all protocols in $\mathcal{S}$, the one that gives the maximum fidelity (denoted as $F_{\mathcal{S}}$) and also find its probability of success (denoted as $p_{\mathcal{S}}$). The same procedure is carried out for the protocol sets $\mathcal{G}$ and $\mathcal{J}$, where the maximum fidelities in $\mathcal{G}$ and $\mathcal{J}$ are denoted as $F_{\mathcal{G}}$ and $F_{\mathcal{J}}$ with corresponding success probabilities $p_{\mathcal{G}}$ and $p_{\mathcal{J}}$. Two searches using the basinhopping algorithm [36] which minimize $F_{\mathcal{G}} - F_{\mathcal{S}}$ and $F_{\mathcal{J}} - F_{\mathcal{S}}$ respectively are carried out within the parameter space $\Sigma$. The input fidelities where our protocols show advantage are where $F_{\mathcal{G}} - F_{\mathcal{S}} < 0$ and $F_{\mathcal{J}} - F_{\mathcal{S}} < 0$. Mathematically, each inequality corresponds to a region in $\Sigma$ and the intersection of the two regions is recorded.

As an example, a point in the intersected region reads $\mathbf{F} = [0.5390, 0.6332, 0.6332, 0.5888]$. When the four input states have those fidelities, the protocol set $\mathcal{S}$ produces a state $(0.6853, 0.0802, 0.0802, 0.1543)^\mathsf{T}$ with fidelity $F_{\mathcal{S}} = 0.6853$ and probability $p_{\mathcal{S}} = 0.2121$. To obtain thus a state, the entangled state $\chi_0$, which has fidelity 0.539, is used to control-SWAP $\chi_1$ and $\chi_2$, which have fidelity 0.6332. Among all protocols in $\mathcal{G}$. the one with a "recurrence" structure first combining $\chi_0$ with $\chi_1$ and $\chi_2$ with $\chi_3$ before combining their purified products yield a state $(0.6842, 0.0553, 0.1314, 0.1291)^\mathsf{T}$ with $F_{\mathcal{G}} = 0.6842$ and $p_{\mathcal{G}} = 0.2069$. Among all protocols in $\mathcal{J}$, the arrangement which first puts $\chi_0$ and $\chi_1$ together for a DEJMPS step followed by a three-pair distillation circuit among its product and $\chi_2$ and $\chi_3$ yields a maximum fidelity $F_{\mathcal{J}} = 0.6842$ and success probability $p_{\mathcal{J}} = 0.2069$ from a state $(0.6842, 0.1314, 0.1291, 0.0553)^\mathsf{T}$. Our set of protocols $\mathcal{S}$ have clear overall advantage over the causally-ordered distillation protocols by producing a state with a higher fidelity and higher success probability than the latter.

As presented in Eq. (39), $\rho_+$ is a weighted mixture of $N_1, N_2, M, T$, and $L$, each of which being an unnormalized mixture of the four Bell states. We calculate the fidelities of all the mixture components of $\rho_+$ and they are: $f(N_1) = f(N_2) = 0.6840$, $f(M) = 0.9746$, $f(T) = 0.3384$, $f(L) = 0.6302$. $M$ is the only component with a fidelity higher than $N_1$ and $N_2$, which are the fidelities of the entangled state produced from doing two definite-ordered DEJMPS steps. This means the fidelity advantage of protocols $\mathcal{S}$ is solely due to the presence of the component $M$. As discussed in Sec. V, the presence of $M$ is due to the nontrivial commutations of the Kraus operators that correspond to the maps of the two distillation steps. This shows that the fidelity advantage of protocols $\mathcal{S}$ indeed comes from two DEJMPS steps being applied in an indefinite causal order.

#### 2. Input states with noise bias

We study the effect of noise bias of the faulty pairs on the distillation fidelities. Noise on the entangled pairs comes from interaction of the particles with environment, during which the target entangled state becomes entangled with the environment. Tracing out the latter leaves the former in a probabilistic mixture of various states. The Pauli-diagonal channel, which results in the mixture components being the target state undergoing Pauli $X$, $Y$, and $Z$ flips, is a fairly complete description of all possible noise models. In practice, the probabilities of undergoing the three flips are different. We define the X-biased channel with a degree $r_X$ as the following CPTP map:

$$\varepsilon_X(\rho) = p \cdot \mathbb{I}\rho\mathbb{I} + r_X(1 - p) \cdot X\rho X$$
$$+ \frac{1 - r_X}{2}(1 - p) \cdot Y\rho Y + \frac{1 - r_X}{2}(1 - p) \cdot Z\rho Z. \tag{45}$$

$r = 1$ indicates a noise channel with complete $X$-bias, $r = 0$ indicates the noise biased away from $X$ and $r = 1/3$ indicates a depolarising channel with equal noise probability, which gives rise to the previously discussed Werner states. A $Y$-biased noise channel with degree $r_Y$ and a Z-biased channel with degree $r_Z$ can be defined in the same manner. Suppose the fidelities of the four faulty pairs are the ones given in the previous discussion ($\mathbf{F} = [0.5390, 0.6332, 0.6332, 0.5888]$) but each faulty pair now has unequal erroneous Bell state components as caused by the noise bias. Figure 6 shows the distillation fidelities $F_{\mathcal{G}}$, $F_{\mathcal{S}}$ and $F_{\mathcal{J}}$ of the three groups of protocols under the three directions of noise bias: (a) for $Y$-biased noise, (b) for $X$-biased, and (c) for $Z$-biased. Protocol $\mathcal{S}$'s fidelity is larger than $\mathcal{G}$'s and $\mathcal{J}$ for relatively unbiased noise model (when $r_X, r_Y$ and $r_Z \approx 1/3$). The fidelity advantage of $\mathcal{S}$ is lost when noise is biased towards or away any of the three directions. We notice in (a) that when noise is biased heavily towards $Y$, using protocols $\mathcal{G}$ results in limited fidelity improvement compared to to $\mathcal{S}$ and $\mathcal{J}$. Similarly in (b), when noise is biased heavily towards $Y$ or $X$, using $\mathcal{J}$ results in small fidelity enhancement. Comparatively, our set of protocols $\mathcal{S}$ result in some amount of fidelity increase under any noise-bias direction. This indicates protocols $\mathcal{S}$ can be advantageous when only fidelities of the input faulty pairs are known but one has little information on the shape of the noise.

### B. Advantageous region of parameters

We have seen that the fidelity advantage of $\mathcal{S}$ exists when input faulty pairs are close to Werner states. In this section, we restrict them to be Werner states and examine the parameter region inside $\Sigma$ where fidelity advantage holds [i.e., $\max(F_{\mathcal{G}} - F_{\mathcal{S}}, F_{\mathcal{J}} - F_{\mathcal{S}}) < 0$]. In general, the region holds a four-dimensional subspace in $\Sigma$. To visualize the region, in Fig. 7, we fix $F_3$ to be equal to (a) 0.5390 and (b) 0.5690 and show the three-dimensional subspace of $[F_0, F_1, F_2]$ bounded by the closed surface. One can see from Fig. 7 that the three-dimensional advantageous region is larger when $F_3 = 0.5390$ than $F_3 = 0.5690$. We have also found (not shown in Fig. 7) that the advantageous region only exists when $F_3 > 0.5$. When $F_3$ is close to 0.5, the region is small. The size of the region

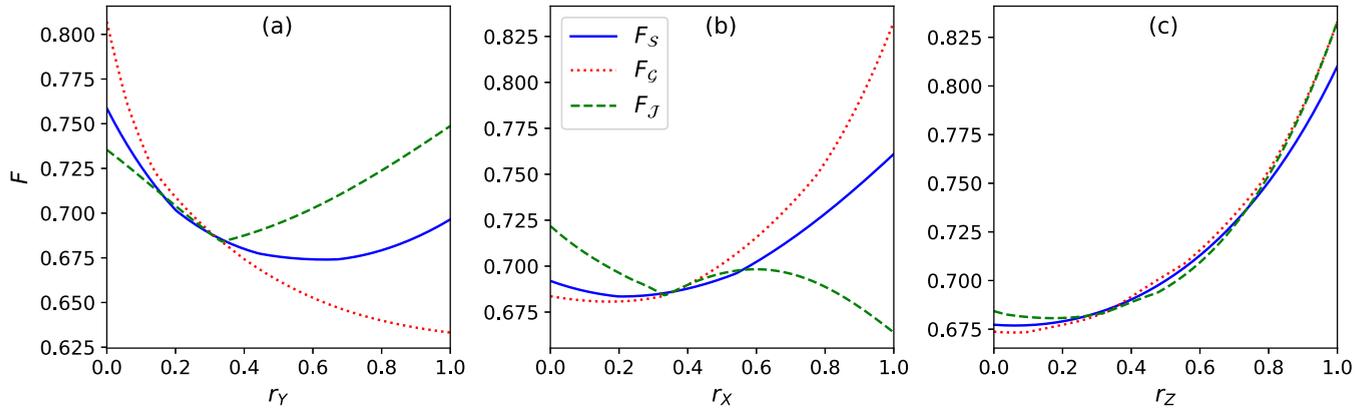

FIG. 6. Distillation fidelities $F_{\mathcal{S}}$(solid blue curve), $F_{\mathcal{G}}$(dotted red curve), and $F_{\mathcal{J}}$ (dashed green curve) for the three groups of protocols $\mathcal{S}$, $\mathcal{G}$ and $\mathcal{J}$ when the four input faulty states have respective fidelities $\mathbf{F} = [0.5390, 0.6332, 0.6332, 0.5888]$ with the same noise bias towards: (a) $Y$(bit and phase) flip, (b) $X$(bit) flip, and (c) $Z$(phase) flip.

increases to $F_3 \approx 0.5390$ after which it decreases to zero when $F_3 \approx 0.75$.

In Fig. 7, the advantageous regions (given by the three-dimensional surfaces) show a discrete three-fold rotational symmetry around the axis $F_1 = F_2 = F_3$. This is expected, as our search algorithm iterates through all protocols within each class of protocols, which include all permutations of the given faulty pairs. As a result, any permutation of given fidelities that are in the advantageous region is also in the advantageous region. One can see from the figures that the regions only exist at some distance away from the axis of symmetry, and that there seem to be two "parts" of the surface: one part lying at lower fidelities and surrounds the symmetry axis, the other part are longer and extends into higher fidelities.

We now present the concrete entanglement distillation protocol within each protocol set that leads to the advantageous region. We choose $F_3 = 0.5390$, $F_2 = 0.5888$. Figures 8(a), 8(b) and 8(c) show the protocols within $\mathcal{G}$, $\mathcal{S}$, and $\mathcal{J}$ that leads to the maximum distillation fidelities $F_{\mathcal{G}}$, $F_{\mathcal{S}}$, and $F_{\mathcal{J}}$ for various $F_0$, $F_1$. The black contours in each plot are the advantageous regions of $\mathcal{S}$, which is essentially a "horizontal slice" of Fig. 8 at $F_2 = 0.5888$. We suggest the reader consult the figure's caption for more information. Here, we point out several features of the graph. In Fig. 8(b), The entirety of each black contour lies within a single region which represents a specific permutation of faulty pairs. As an example, the previously presented $\mathbf{F} = [0.6332, 0.6332, 0.5888, 0.5390]$ which has $(F_0, F_1) = (0.6332, 0.6332)$ and belongs to the contour in the top-right corner has a permutation "$(0,1,2,3)$", which uses entangled pair 3 (with a fidelity $F_3 = 0.5390$) to C-SWAP the zeroth and first entangled pairs (with fidelities 0.6332). Interestingly, we find in Fig. 8(b) that it is always the entangled pair with the lowest fidelity being used to C-SWAP the two highest-fidelity pairs that lead to the highest distilled fidelity in $\mathcal{S}$. Second, one can also see from Figs. 8(a) and 8(c) that the contours of advantage all lie at the boundaries and points of intersections of different regions of $\mathcal{G}$ and $\mathcal{S}$'s protocols. The reason for this is not known to us. Third, in (a) and (c), the contour in the top-right corner overlaps with region **P**, which only uses the zeroth (with fidelity 0.6332), first (with fidelity 0.6332) and second faulty pairs (with fidelity 0.5888) and discards the third pair (with the lowest fidelity 0.5390). In contrast, our protocols $\mathcal{S}$ do not discard the lowest-fidelity pair but use it to further enhance the fidelity of distillation. This suggests our protocols may use the entangled pairs more efficiently, which is beneficial when rate of entanglement distribution is low.

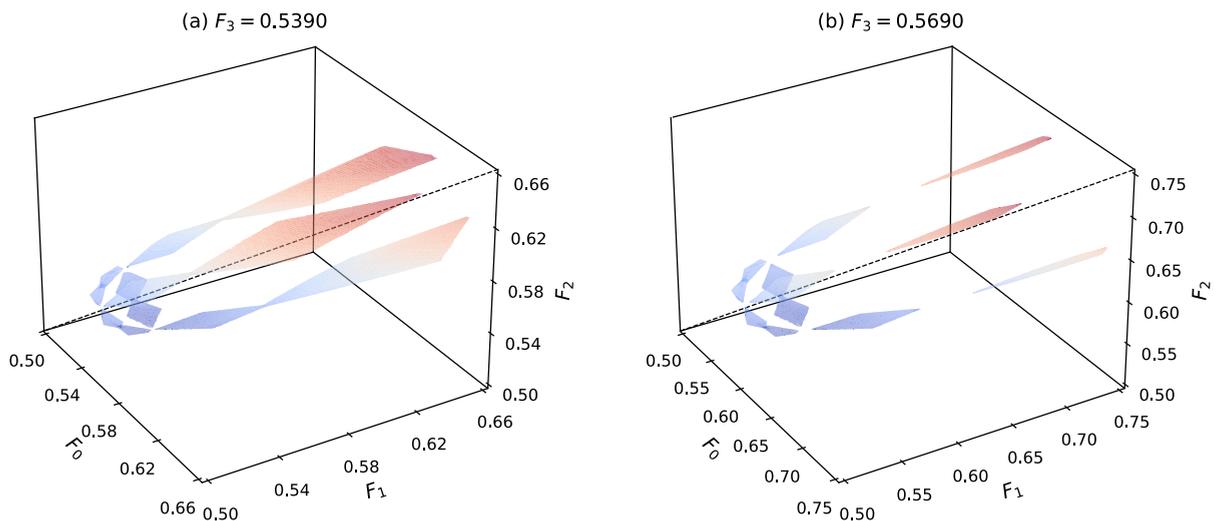

FIG. 7. Regions of $[F_0, F_1, F_2]$ with (a) $F_3 = 0.5390$ and (b) 0.5690 where protocols $\mathcal{S}$ have higher fidelity than $\mathcal{G}$ and $\mathcal{J}$ [i.e., $\max(F_{\mathcal{G}} - F_{\mathcal{S}}, F_{\mathcal{J}} - F_{\mathcal{S}}) < 0$]. The dashed line is specified by $F_0 = F_1 = F_2$.

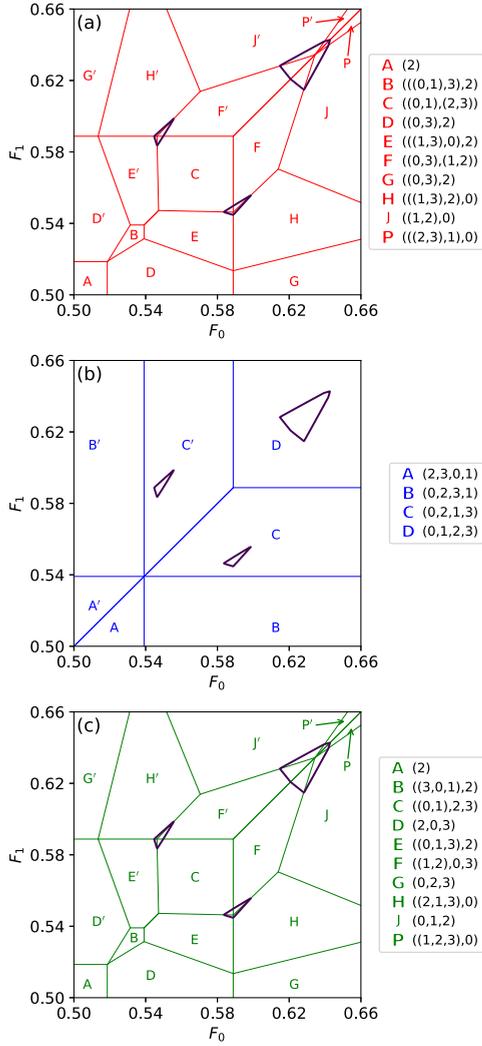

FIG. 8. Concrete distillation protocols within each set of protocols (a) $\mathcal{G}$, (b) $\mathcal{S}$, (c) $\mathcal{J}$ that lead to maximum fidelities for four Werner states with various $F_0$ and $F_1$ when $F_3 = 0.5390$ and $F_2 = 0.5888$. Within the black contours are where protocol $\mathcal{S}$ has fidelity advantage over $\mathcal{G}$ and $\mathcal{J}$. Meaning of notations in the legends that denote the protocols are given as below. (a) A single number $i$ in a pair of parentheses means the $i$th entangled pair (with fidelity $F_i$) is simply taken as the output and all other pairs are discarded. Two numbers $i$ and $j$ in a pair of parentheses means the $i$th and $j$th entangled pairs are DEJMPS-distilled. If there is an outer pair of parentheses present, the distillation product of the inner pair of parentheses is used as the input to the DEJMPS step described by the outer pair. (b) "$(i, j, k, l)$" denotes the C-SWAP protocol where $l$th entangled pair C-SWAPs the $i$th and $j$th entangled pairs. (c) Two numbers in a pair of parentheses mean those two entangled pair DEJMPS distilled. Three numbers in a pair of parentheses mean those pairs are distilled using the three-pair distillation circuit as given in Fig. 2 with the order of appearance before the circuit the same as the order with which the corresponding number appears in the parentheses. If an outer pair of parentheses encompass an inner pair, the distillation product of the inner parentheses is used as the input to the distillation step described by the outer parentheses. A letter with a prime represents a distillation protocol denoted with "0" and "1" swapped compared with the letter without prime. For example, **F'** in (a) represents a protocol denoted by $((1,3),(0,2))$ whereas F denotes the protocol $((0,3),(1,2))$.

## VII. DISCUSSIONS

We have presented and studied the practical benefit of applying indefinite causal order in the task of entanglement distillation, which is an important and necessary protocol in practical quantum communication. When four faulty entangled pairs subject to Pauli noise are shared, we have constructed a protocol where one faulty pair is used to control-SWAP two other faulty pairs before two steps of the basic DEJMPS entanglement distillation protocol are applied onto a fourth faulty pair and the two SWAP-ed pairs. We have shown that the constructed protocol can be seen as applying two DEJMPS steps in a superposition of two causal orders. This is done by showing that the overall trace-decreasing map of the protocol can be expressed in the same form as a quantum switch which indicates the presence of indefinite orders of two trace-decreasing maps of the constituent DEJMPS steps. It is also shown that for some input faulty states, the protocol returns an output entangled state with higher fidelity and success probability than a wide range of protocols constructed from concatenation of smaller entanglement distillation protocols that follow a definite causal order. This includes concatenation of multiple DEJMPS steps, and concatenation between a DEJMPS step and a typical three-pair distillation protocol constructed from three-bit stabilizer quantum error-detecting code. The circuit of our protocol has relatively low complexity, making itself viable to implement and demonstrate experimentally.

We believe effort should be made into understanding whether, at least for the examples we presented in this paper, the advantage of fidelity/probability of success is really due to indefinite causal order *per se*, or can it be replicated/exceeded with definite causal order protocols. This is to complement the recent debate on indefinite-causal-order advantage in other application settings. For example, the effect of noise reduction from putting two noisy Pauli channels in an indefinite causal order as presented in Ref. [11] was later matched in Ref. [37] with a setup (Fig. 6 of the paper) which consists of the two noisy channels arranged in a definite causal order, but there is an extra noiseless side channel generated from the control state via an entangling gate between the control and target states. The authors of Ref. [37] argue such a setup shows that Ref. [11]'s noise advantage is not due to indefinite causal order per se because it can be realized with other kinds of causally-ordered resources. Here in our practical setup, the meaning of "resource" and what count as "free resource" is different and more specific. In the task of entanglement distillation, local unitary gates and classical communication are seen as free resource, and faulty entanglements are not free (as they are hard to generate). The quantum switch in our circuit should be seen as free resource as it is realized purely with local unitary gates. In comparison, other "free resources" that feature "definite causal order" in our setting are other types of entanglement distillation protocols outside of the sets $\mathcal{G}$ and $\mathcal{J}$. These protocols can be built from scratch using local unitary gates. Another piece of work that questioned the sole-advantage of quantum switch in noise reduction is [38]. The authors proposed a setup which simply puts each of the two channels on two separate paths and let the photon propagate through a superposition of the two paths. For the task of entanglement distillation, such a spirit can be realized as the following protocol: using a faulty Bell pair (which corresponds to the control qubit in Ref. [38]) to C-SWAP two other faulty pairs before a distillation step is carried out between a fourth faulty pair and one of the SWAPed pairs, followed by a measurement on the control pair. This simulates a process in which the fourth pair is distilled with a superposition of either one of the two swapped pairs followed by a postselection via measuring the control pair. Compared with protocol $\mathcal{S}$'s circuit in Fig. 5, one can see that this is simply a sub-routine of that circuit which does not have the second DEJMPS step at the end. This indicates it is unlikely to surpass the performance of the our protocols $\mathcal{S}$.

On the other hand, there are numerous possible extensions of our work. A natural and immediate one is to compare the advantage of fidelity and success probability of applying $> 2$ DEJMPS steps in an indefinite causal order over definite-causal-order protocols that distill the same number of faulty pairs. This will require multiple Bell states, or a smaller number of higher-dimensional entangled states to act as the control state of the quantum switch. Alternatively, one may also consider still using one faulty Bell state as the control state by keeping the number of constituent CP maps in the superposition at two where each constituent CP map will be an entanglement distillation protocol that uses more faulty pairs. Extensions to superposing the causal orders of multipartite entanglement distillation protocols and one-way entanglement distillation protocols can also be carried out. Given the known connection [22,31] of one-way entanglement distillation protocols with stabilizer quantum error correction codes, we hope the possible existence of advantage can stimulate effort into incorporating indefinite causal structures into the encoding/decoding of quantum error-correcting codes, which is a vital part of practical quantum information processing. Additionally, modifications similar to our proposal can be envisioned for many other distillation-like and breeding-like protocols that feature repeated applications of some subroutine, each featuring a subsystem interacting with ancillary states. Some examples include: distillation of magic states for universal quantum computation [39,40], and repeated breeding of oscillator states [41] for distillation of bosonic quantum error-correcting codewords.

## ACKNOWLEDGMENTS

We acknowledge financial supports from Imperial College London through the Presidential Scholarship, the EPSRC through EP/T00097X/1 and EP/T001062/1 and the Korean Institute of Science and Technology for their Open Innovation Lab programme.

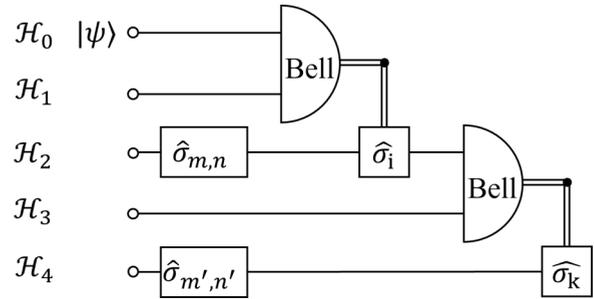

FIG. 9. Circuit for two concatenated steps of noisy teleportation. $\mathcal{H}_i$, where $i \in \{0, 1, 2, 3, 4\}$ denotes the Hilbert spaces of the corresponding qubits.

## APPENDIX: TWO QUANTUM TELEPORTATION STEPS APPLIED UNDER INDEFINITE CAUSAL ORDER

One noticeable proposal for application of indefinite causal order in quantum communication is putting two teleportation steps in a superposition of causal orders. The standard quantum teleportation protocol [42] has a $|\Phi^+\rangle$ state shared between Alice and Bob. For Alice to teleport her qubit $|\psi\rangle$, she performs a Bell measurement between $|\psi\rangle$ and her qubit of $|\Phi^+\rangle$. The measurement result is sent to Bob via a classical channel. Bob maps the measurement result onto a Pauli operator which he performs on the teleported state to recover $|\psi\rangle$. It is known from [43] that when the shared entangled state is noisy, the above protocol is essentially $|\psi\rangle$ undergoing a generalized depolarising channel. Two noisy teleportation steps applied in series naturally degrades $|\psi\rangle$ more than one step. It was claimed, however, in Refs. [18–20] that applying two teleportation steps in an indefinite causal order reduces the noise of the final target state as compared to the two steps applied back-to-back. In Ref. [19], a photonic implementation of this scheme is proposed where entangled photon pairs pass through beam splitters and subsequent Bell measurements such that a distinct causal order is featured on each output path of the beam splitters. On one of the two paths, SWAP gates are performed on the photon pairs to simulate the swapping of teleportation channels. The author in Ref. [19] considered the case when the entangled pair is pure but not maximally entangled. In this section, we examine their proposals [18–20] more carefully and show that no noise reduction of the target state actually occurs.

We first consider two concatenated standard teleportation steps with the input target state $|\psi\rangle\langle\psi|$ in Hilbert space $\mathcal{H}_0$, whose circuit is shown in Fig. 9. A pure but not maximally entangled state $\chi$ is in Hilbert space $\mathcal{H}_1 \otimes \mathcal{H}_2$:

$$\chi = \sum_{mn} \sigma_m^{(2)} |\Phi^+\rangle\langle\Phi^+| \sigma_n^{(2)} q_{mn} \tag{A1}$$

where $\sigma_m$ and $\sigma_n$ are Pauli operators, the superscript "2" means the Pauli noise occurs on the qubit of $\chi$ in Hilbert space $\mathcal{H}_2$ as denoted in Fig. 9. $q_{mn}$ denotes the entries of the density matrix under the Bell basis. A second pure yet not maximally entangled pair $\xi$ lies in Hilbert space $\mathcal{H}_3 \otimes \mathcal{H}_4$:

$$\xi = \sum_{m'n'} \sigma_{m'}^{(4)} |\Phi^+\rangle\langle\Phi^+| \sigma_{n'}^{(4)} s_{m'n'}. \tag{A2}$$

The superscripts "4" again denotes the Hilbert space on which we assume Pauli noise occurs. The first teleportation step carried out with $\chi$ and the original target state $|\psi\rangle\langle\psi|$ yields

$$\rho = \sum_{nmi} \sigma_i^{(2)} G_i^{(01)} |\psi\rangle\langle\psi| \otimes \sigma_n^{(4)} |\Phi^+\rangle\langle\Phi^+| \sigma_m^{(4)} q_{mn} G_i^{(01)} \sigma_i^{(2)} \tag{A3}$$

where $G_i^{(01)}$ is the Bell measurement projector associated with outcome $i$ and $\sigma_i^{(2)}$ is the corresponding Pauli correction applied on the output Hilbert space. We now want to express the errors on $\rho$ in terms of the errors $\sigma_m$, $\sigma_n$ on $\chi$. The Pauli errors $\sigma_n$ and $\sigma_m$ commute with $G_i^{(01)}$ as they act on different Hilbert spaces. They commute with $\sigma_i$ if $m(n) = i$, or anti-commute with $\sigma_i$ if otherwise. This means

$$\begin{aligned}\rho &= \sum_{nmi} \sigma_n^{(2)} \left( \sigma_i^{(2)} G_i^{(01)} |\psi\rangle\langle\psi| \otimes |\Phi^+\rangle\langle\Phi^+| G_i^{(01)} \sigma_i^{(2)} \right) \\ &\quad \sigma_m^{(2)} A_{ni} A_{im} q_{mn} \\ &= \sum_{nmi} \sigma_n^{(2)} |\psi\rangle\langle\psi| \sigma_m^{(2)} A_{ni} A_{im} q_{mn}, \end{aligned} \tag{A4}$$

where $A$ is a global phase caused by permuting $\sigma_{m,n}$ with $\sigma_i$. $A_{ni} = 1$ when $[\sigma_n, \sigma_i] = 0$ and $= -1$ if they anti-commute. The second line is due to the part in the large parenthesis is simply $|\psi\rangle\langle\psi|$ which comes from a perfect teleportation with noiseless $|\Phi^+\rangle$. We note that $\sum_i A_{ni} A_{im} = \delta_{nm}$. This eliminates all summing components where $n \neq m$ and we arrive at

$$\rho = \sum_m \sigma_m |\psi\rangle\langle\psi| \sigma_m q_{mm}, \tag{A5}$$

which, as expected from [43], is the original state $|\psi\rangle$ undergone a depolarising channel. $\rho$ now goes through the second teleportation channel which has an output $\rho'$ expressed as

$$\begin{aligned}\rho' = \sum_{mm'n'k} &\sigma_k^{(4)} G_k^{(23)} \left[ \sigma_m^{(2)} |\psi\rangle\langle\psi| \sigma_m^{(2)} q_{mn} \right] \otimes \\ &\times \sigma_{m'}^{(4)} |\Phi^+\rangle\langle\Phi^+| \sigma_{n'}^{(4)} s_{m'n'} G_k^{(23)} \sigma_k^{(4)}, \end{aligned} \tag{A6}$$

where the superscripts denote the Hilbert spaces of the corresponding operations. Like before, we move $\sigma_{m'}^{(4)}$ and $\sigma_{n'}^{(4)}$ across the Pauli corrections during which global phases $A_{m'k}$ and $A_{n'k}$ arise:

$$\begin{aligned}\rho' = \sum_{mm'n'k} &\sigma_{m'}^{(4)} \sigma_k^{(4)} G_k^{(23)} \left[ \sigma_m^{(2)} |\psi\rangle\langle\psi| \sigma_m^{(2)} q_{mn} \right] \otimes |\Phi^+\rangle\langle\Phi^+| \\ &\times s_{m'n'} G_k^{(23)} \sigma_k^{(4)} \sigma_{n'}^{(4)} A_{m'k} A_{n'k}. \end{aligned} \tag{A7}$$

We notice that $\rho$ now can be interpreted as the inner state $\sigma_m^{(2)} |\psi\rangle\langle\psi| \sigma_m^{(2)}$ passing through a perfect teleportation channel followed by $\sigma_{m'}^{(4)}$ on the ket side (or $\sigma_{n'}^{(4)}$ on the bra side). Since a perfect teleportation preserves the target state, a Pauli error $\sigma_m^{(2)}$ *commutes* with a perfect teleportation. Having an error on the input target state has the same effect as having the error on the perfectly teleported state. We can hence move $\sigma_m^{(2)}$ across $\sigma_k^{(4)}$ and $G_k^{(23)}$, giving

$$\begin{aligned}\rho' &= \sum_{mm'n'k} \sigma_{m'}^{(4)} \sigma_m^{(4)} \left[ \sigma_k^{(4)} G_k^{(23)} |\psi\rangle\langle\psi| \otimes |\Phi^+\rangle\langle\Phi^+| G_k^{(23)} \sigma_k^{(4)} \right] \\ &\quad \times \sigma_m^{(4)} \sigma_{n'}^{(4)} A_{m'k} A_{n'k} q_{mn} s_{m'n'} \\ &= \sum_{mm'n'k} \sigma_{m'}^{(4)} \sigma_m^{(4)} |\psi\rangle\langle\psi| \sigma_m^{(4)} \sigma_{n'}^{(4)} A_{m'k} A_{n'k} q_{mn} s_{m'n'} \\ &= \sum_{mm'n'k} \sigma_{m'}^{(4)} \sigma_m^{(4)} |\psi\rangle\langle\psi| \sigma_m^{(4)} \sigma_{n'}^{(4)} \delta_{m'n'} q_{mn} s_{m'n'} \\ &= \sum_{mn} \sigma_n \sigma_m |\psi\rangle\langle\psi| \sigma_m \sigma_n q_{mn} s_{nn}. \end{aligned} \tag{A8}$$

We now calculate the state produced from the scheme proposed in [19] which uses a control qubit $|c\rangle = |+\rangle = 1/\sqrt{2}(|0\rangle + |1\rangle)$ to swap $\chi$ and $\xi$, then does the two teleportation steps before measuring $|c\rangle$ in the Fourier basis and postselecting the "+" outcome. Before measuring $|c\rangle$, the overall state $\hat{R}$ which consists of $|c\rangle$ and teleportation target state $|\psi\rangle$ reads

$$\begin{aligned}R ={}& \frac{|0\rangle\langle 0|}{2} \otimes \sum_{mn} \sigma_n \sigma_m |\psi\rangle\langle\psi| \sigma_m \sigma_n q_{mm} s_{nn} + \frac{|1\rangle\langle 1|}{2} \otimes \sum_{mn} \sigma_m \sigma_n |\psi\rangle\langle\psi| \sigma_n \sigma_m q_{mm} s_{nn} \\ &+ \frac{|0\rangle\langle 1|}{2} \otimes \sum_{mnm'n'ik} \sigma_k^{(4)} G_k^{(23)} \left[ \sigma_i^{(2)} G_i^{(01)} |\psi\rangle\langle\psi| \otimes \sigma_m^{(2)} |\Phi^+\rangle\langle\Phi^+| \sigma_{n'}^{(2)} q_{mn} G_i^{(01)} \sigma_i^{(2)} \right] \otimes \sigma_{m'}^{(4)} |\Phi^+\rangle\langle\Phi^+| \sigma_n^{(2)} \\ &\times s_{m'n'} G_k^{(23)} \sigma_k^{(4)} + \frac{|1\rangle\langle 0|}{2} \otimes \sum_{mnm'n'ik} \sigma_k^{(4)} G_k^{(23)} \left[ \sigma_i^{(2)} G_i^{(01)} |\psi\rangle\langle\psi| \otimes \right. \\ &\times \left. \sigma_m^{(2)} |\Phi^+\rangle\langle\Phi^+| \sigma_n^{(2)} q_{mn} G_i^{(01)} \sigma_i^{(2)} \right] \otimes \sigma_m^{(4)} |\Phi^+\rangle\langle\Phi^+| \sigma_{n'}^{(2)} s_{m'n'} G_k^{(23)} \sigma_k^{(4)}. \end{aligned} \tag{A9}$$

For the last two terms in the summation, we follow a similar strategy by moving the Pauli errors on the entangled pairs pass the Pauli corrections where global phases arise from commutation (or anticommutation). This gives

$$\begin{aligned}R ={}& \frac{|0\rangle\langle 0|}{2} \otimes \sum_{mn} \sigma_n \sigma_m |\psi\rangle\langle\psi| \sigma_m \sigma_n q_{mm} s_{nn} + \frac{|1\rangle\langle 1|}{2} \otimes \sum_{mn} \sigma_m \sigma_n |\psi\rangle\langle\psi| \sigma_n \sigma_m q_{mm} s_{nn} + \frac{|0\rangle\langle 1|}{2} \otimes \sum_{mn} \sigma_n \sigma_m |\psi\rangle\langle\psi| \sigma_m \sigma_n q_{mn} s_{nm} \\ &+ \frac{|1\rangle\langle 0|}{2} \otimes \sum_{mn} \sigma_m \sigma_n |\psi\rangle\langle\psi| \sigma_n \sigma_m q_{mn} s_{nm}. \end{aligned} \tag{A10}$$

The fact that $\chi$ and $\xi$ are pure states means that $s_{mn}$ and $q_{mn}$ are factorizable: they can be written as $s_{mn} = u_m u_n^*$ and $q_{mn} = v_m v_n^*$ where $\mathbf{u} = \{u_m\}$ and $\mathbf{v} = \{v_m\}$ are normalized probability amplitudes of each Bell component such that $|\mathbf{u}| = |\mathbf{v}| = 1$. We consider a common situation (which is also the case considered in Ref. [19]) where the two faulty pairs, $\chi$ and $\xi$, are identical. This is motivated from the speculation that they may come from the same single photon generator. This means $\mathbf{u} = \mathbf{v}e^{i\phi}$ with some constant phase factor $\phi$. We then have $q_{mn}s_{nm} = v_m v_n^* u_n u_m^* = v_m(u_n^* e^{i\phi})u_n$ $(v_m^* e^{-i\phi}) = v_m v_n^* u_n u_m^* = q_{mm}s_{nn}$. Substituting this into Eq. (A10) and notice that $\sigma_n \sigma_m |\psi\rangle\langle\psi| \sigma_m \sigma_n = \sigma_m \sigma_n |\psi\rangle\langle\psi| \sigma_n \sigma_m$ for any $n, m$ since $\sigma_n$ and $\sigma_m$ either commute or anticommute, one can express $\hat{R}$ as

$$R = |+\rangle\langle+| \otimes \rho', \tag{A11}$$

which yields an output state $\rho'$, the same as that from two definite-order teleportation steps, regardless of the basis and outcome of measurement on the control qubit. no noise reduction has occurred.

We leave as future work to examine the more general case when $\chi$ and $\xi$ are mixed states. We expect, however, more practical challenges to implement the scheme in Ref. [19] in this case. This is because the mixed Pauli noise is likely to occur during storage or transmission of the entangled states. It is not hard to see that in order for noise interference to occur due to indefinite causal order, controlled-swapping of the two entangled pairs must happen after, not before the mixed Pauli noises. This means if, for example, the major source of noise comes from the physical communication channel during transmissions of the entangled pairs, the controlled-swap will have to be carried out between remote parties. Suppose the two faulty entangled pairs are shared between Alice & Bob and Bob & Charlie, respectively, then extra perfect Bell pairs will have to be pre-shared between Alice & Charlie for the remote-swap. The required extra resources bring additional challenges to the practical implementation.

---